\documentclass[amsmath, floats,floatfix, superscriptaddress, nofootinbib, showkeys]{revtex4}
\usepackage{epstopdf}
\usepackage{pdfpages}
\usepackage{subfigure}
\usepackage{amsmath}
\usepackage{eqnarray,amsmath}
\usepackage{amssymb}
\usepackage{amsthm}
\usepackage{epstopdf}
\usepackage{mathrsfs}
\usepackage{natbib}
\usepackage{amssymb,latexsym}
\usepackage{dcolumn}
\usepackage{graphicx}
\usepackage{multirow}
\def\be{\begin{equation}}
	\def\ee{\end{equation}}
\def\ba{\begin{eqnarray}}
	\def\ea{\end{eqnarray}}

\begin{document}
	\title{Radiation from Einstein-Gauss-Bonnet de Sitter Black Hole via Tunneling Process}
	
\author{Sareh Eslamzadeh}
\email{S.Eslamzadeh@stu.umz.ac.ir}
\affiliation{Department of Theoretical Physics, Faculty of Basic Sciences, University of Mazandaran, P. O. Box 47416-95447, Babolsar, IRAN}
\affiliation{ICRANet-Mazandaran, University of Mazandaran, P. O. Box 47416-95447, Babolsar, IRAN}
	
\author{Javad T. Firouzjaee}
\email{firouzjaee@kntu.ac.ir}
\affiliation{Department of Physics, K. N. Toosi University of Technology, P.O. Box 15875-4416, Tehran, Iran}
\affiliation{School of Physics, Institute for Research in Fundamental Sciences (IPM), P.O. Box 19395-5531, Tehran, Iran}
		
\author{Kourosh Nozari}
\email{knozari@umz.ac.ir}
\affiliation{Department of Theoretical Physics, Faculty of Basic Sciences,
		University of Mazandaran, P. O. Box 47416-95447, Babolsar, IRAN}
\affiliation{ICRANet-Mazandaran, University of Mazandaran, P. O. Box 47416-95447, Babolsar, IRAN}

	\begin{abstract}
		\textbf{Abstract:} \\ 		
		In this paper, we probe in 4D Einstein-Gauss-Bonnet black hole and its thermodynamics. We illustrate the three asymptotically 4D EGB spacetime as an asymptotically flat, de Sitter, and Anti-de Sitter. Also, we apply the tunneling of the massless particles from the horizon of 4D EGB gravity and we investigate the correlation between the emission modes and temperature of the horizon. In asymptotically flat spacetime, the existence of the coupling constant alone constructs the regular spacetime, the radiation deviates from the pure thermal, and the temperature of the black hole horizon would be zero in the final stage of the black hole evaporation.  In Asymptotically de Sitter spacetime, results illustrate that the evolution of the temperatures is in direction of the remaining rest mass with the probably same temperature for the black hole and the cosmological horizon. In addition, the exciting result is that temperature behaviors exactly match with the temperature behaviors of a regular black hole in Lovelock gravity in a higher dimension. 
		
		
	\end{abstract}
	%
	%
	
	\maketitle
	
	\tableofcontents
	\newpage
	\section{Introduction}
	The theory of GR, as the most powerful theory in mapping spacetime, still suffers from an unresolved mystery.  The most important of these mysteries that GR predicts is the behavior of spacetime in distances of the order of the Planck scales and singularities. Among theories that correct the GR, we pay special attention to the category that modifies gravity and answers the question: Is gravity always attractive?\
	
	From GR, we know that if $K_{\mu\nu}$ be a rank-2 tensor related to the curvature of spacetime and responsible for the geometry, then $(i)$ this is necessarily constructed of metric tensors and curvature tensor (first two derivatives of the metric tensor). Also, since the stress-momentum tensor is divergence-free then $(ii)$ $K_{\mu\nu}$ should also be divergence-free. Moreover, since the stress-momentum tensor is symmetric then $(iii)$ $K_{\mu\nu}$ should also be symmetric; And, $(iv)$ $K_{\mu\nu}$ should not contain terms higher than linear in the second-order-derivatives of the metric tensor. Already, it has been shown \cite{Car922,WV2217}, to maintain these constraints, the only possible form for $K_{\mu\nu}$ has the form $K_{\mu\nu}=a G_{\mu\nu}+b g_{\mu\nu}$. In 1971, Lovelock has shown, considering higher dimensions, the third and fourth restrictions are excess \cite{Lov71}. Indeed, Lovelock's theorem is a natural higher curvature generalization of Einstein's gravity. The first Lovelock term is included the Einstein-Hilbert action and its second term is known as the Gauss-Bonnet term. Among the modified gravity theories, Einstein-Gauss-Bonnet theory has an important role in the higher dimensional generalization of Einstein's theory. Two important issues make this interesting: First, if one works at tree level in the string coupling $\alpha'$ and performs a perturbative expansion in the inverse string tension, one will see that the leading order term gives the standard Einstein-Hilbert action, and the other leading order term gives the Gauss-Bonnet action, as regards the string theory known to be ghost-free, such a result is expected \cite{GrSl87, MeTs87}. Another interesting issue is the Einstein-Gauss-Bonnet theory keeps the second-order equation of motion in an arbitrary number of dimensions $D\ge5$. This important feature provides the basis for research on gravity in the other framework. Recently, Glavan and Lin, by presenting an intelligent method, derived the 4-dimensional EGB gravity by rescaling the coupling constant \cite{GlLi20}. In a heterotic string theory, the coupling constant, $\alpha$, is associated with the slope parameter, $\alpha'$, and has dimensions of $(length)^2$. Also, we should pay attention to consistency, if we do identify EGB gravity with the stringy generalization of GR then we must restrict attention to $\alpha>0$ \cite{KlFr12}. Authors in Ref. \cite{GlLi20} show that if we multiply the GB action by a factor $\frac{\alpha}{D-4}$ instead of $\alpha$, then we will take the non-vanishing contribution to Einstein's equations in $D=4$. Besides, some authors commented on this method to drive 4D EGB gravity\cite{GCT20,HKMP20,Koba20,FCCM20,LP20,AGM20}. In Refs. \cite{HKMP20, FCCM20, Koba20, LP20}, the authors showed that the well-defined version of the 4D EGB theory is a special scalar-tensor theory described by a subclass of the Horndeski theory. For instance, the authors in Ref. \cite{FCCM20} worked on introducing a counter term into the action without considering any extra dimensions and discussed in their theory and the original version of it. Also, in Ref. \cite{Koba20}, the author proposed the scalar-tensor reformulation of the regularized version of Lovelock gravity in four dimensions. Furthermore, in Ref. \cite{AGM20}, Aoki, Gorji, and Mukayama proposed the consistent 4D EGB theory. They looked into the gravitational degrees of freedom and the diffeomorphism invariance in the original and consistent theories.\
	
	Anyway, the 4D EGB gravity has been attracted attention in a broad context of the black hole and its thermodynamics articles. The observational constraints on regular 4D EGB gravity and information of the range of coupling constant have been searched in Ref. \cite{CliFer20}.  Instability, radiation, and the grey-body factors, quasinormal modes, and shadows of black holes in novel 4D EGB gravity have been investigated in Refs. \cite{Sho20,ZhLi20,KoZi20,KoZin20,GhMa20,KoZh20}. In Ref. \cite{ZhLi20}, both the greybody factor and the power spectra of the Hawking radiation of massless scalar have been studied. Also, the dependence of the various temperatures of the black hole versus the coupling constant and the cosmological constant has been illustrated. In Refs. \cite{KoZi20, KoZin20}, graybody factors and corresponding energy emission rates for Dirac, electromagnetic and gravitational fields in asymptotically flat 4D EGB theory have been investigated and the lifetime of the black hole for various values of the coupling constant has been estimated. In Ref. \cite{HHX21}, the evaporation process of AdS EGB black holes in D-dimensional cases with both positive and negative coupling constant has been studied. Moreover, some research has been done about charged black holes in 4D EGB gravity. For example, In Refs. \cite{PFer20,EsJa20,GhMa21}, charged black hole in AdS 4D EGB spacetime has been studied and shadow, energy emission, and heat engine have been calculated. In Ref. \cite{GhMa21}, horizons and extremality, temperature, heat capacity, and phase transition have been investigated. In Ref. \cite{WL20}, thermodynamics and phase transition of the 4D EGB charged black hole has been researched. Also, temperature and heat capacity, free energy, shadow, and quasinormal modes of charged black holes in 4D EGB gravity has been researched in Refs. \cite{Jus20,JaKo21,SiGh20}. In Ref. \cite{YWCW20}, the motion of charged test particle in the novel charged 4D EGB black hole have been considered and the effect of the GB coupling constant on the validity of the weak cosmic censorship conjecture has been investigated.\
	
Semiclassical methods of modeling Hawking radiation as a tunneling effect were developed over the past two decades and have generated much interest. This approach has been introduced by Hawking when he proposed his exciting theoretical discovery named "Hawking radiation" \cite{Haw74} and after that, it has been explained by Parikh and Wilczek \cite{Wil00, Par04} that how Hawking radiation happens. Hawking explained that because of quantum fluctuations of the vacuum in the neighborhood of the black hole horizon, the particle and antiparticle create; then, a created particle can tunnel through the horizon and a distant observer can detect it as radiation of black hole. Parikh and Wilczek completed this theoretical picture based on the WKB approximation. This approximation uses geometric optic approximation which is another aspect of eikonal approximation in wave interpretation \cite{Firouzjaee:2014zfa}. Particles are placed in front of a barrier in the height of their energy, then they tunnel from this barrier. After particle tunneling, the mass of the black hole is reduces by particles' energy. Although a more accurate picture of the tunneling was provided with the Parikh-Wilczek approach, there were still unsolved problems such as temperature divergence, unitary, and information loss.  Many researches have been done on the radiation and tunneling process from the various black hole horizons; Some of these most important researches can be seen in Refs. \cite{MVag05,Kra95,Arz06,AkhSin06,BaMaj08,ZhCa08,ArzMed05,NozMe08,CriVa08,KerMa08,HeCri08,FirMan12,Med02,Firouzjaee:2014zfa,EsNoz20}. We apply the Parikh-Wilczek method and will see that how considering dS 4D EGB spacetime can solve these problems. Finally, to complete the discussion of tunneling methods, it is good to mention that the approach in which a spherical shell replaces the particle and this shell of matter interacts with a black hole \cite{Kra95} or the approach based on the complex paths \cite{99Pad,02Pad}.\\

In this paper, we focus on the thermodynamics of the spherically symmetric 4D EGB black holes with asymptotically flat, de Sitter, and Anti de Sitter spacetimes. We will apply the tunneling of the massless particles from the horizon of 4D EGB black holes and we will investigate the correlation between the emission modes and temperature of this horizon. This paper is organized as follows: In section II, we illustrate the spacetime of the 4D EGB in three asymptotical situations: an asymptotically flat, de Sitter, and Anti de Sitter spacetime. In section III, we explain the tunneling process analytically and illustrate the temperature of the 4D EGB black hole in asymptotically flat and asymptotically de Sitter spacetimes. Section IV contains a summary and conclusion.\\
	
	\section{Structure of 4D EGB Black Holes}
	In this section, we analyze the EGB gravity and look into it from asymptotically flat, dS, and AdS spacetime. The action of EGB gravity in D-dimensional spacetime has the form
	
	\begin{equation}
		S_{EGB}=S_{EH}+S_{GB}=\frac{1}{16\pi} \int d^{D}x \sqrt{-g}\: [R-2 \Lambda +\alpha \mathcal{L}_{GB}],
	\end{equation}
	\\
	where $R$ is the Ricci scalar, $\Lambda$ is the cosmological constant, $\alpha$ is a positive dimensionless constant and $\mathcal{L}_{GB}$ is the Gauss-Bonnet invariant as follows
	
	\begin{equation}
		\mathcal{L}_{GB}=R^{\mu \nu}_{\rho \sigma} R^{\rho \sigma}_{\mu \nu}-4 R^{\mu}_{\nu}R^{\nu}_{\mu}+R^2.
	\end{equation}
	
	Recently, Glavan and Lin investigated the rescaling of $\alpha \to \frac{\alpha}{D-4}$ and discovered a novel black hole solution as the 4D EGB solution \cite{GlLi20}. Of course, we should emphasize that their result must be reconsidered based on the consistent theory \cite{GCT20,HKMP20,Koba20,FCCM20,LP20, AGM20}. In the direct approach, one can start from the action of the consistent theory and solve the corresponding equations of motion \cite{GCT20,HKMP20,LP20}. In the indirect approach, one can check whether the Weyl tensor for the $(D-1)$-Dimensional spatial part of the D-dimensional counterpart of the solution vanishes or not. If it vanishes, the solution is also a solution of the consistent theory \cite{AGM20}.
	If  the general form of the metric has been considered as the form
	
	\begin{equation}
		ds^2=-g_{00}(r) dt^2+g_{00}^{-1}(r) dr^2+ r^2 d\Omega^2,
	\end{equation}
	\\
	with the explanations above, there is the spherically symmetric solution of EGB spacetime in four-dimensional as follows
	
	\begin{equation}\label{eqg00}
		g_{00}(r)=1+\frac{r^{2}}{2\alpha} \Big[1-\sqrt{1+\frac{8 M \alpha}{r^{3}}+\frac{4 \alpha \Lambda}{3}}\Big].
	\end{equation}
	
	This solution constructs three asymptotically spacetime that we probe these in the following subsections.\\

	\subsection{Asymptotically Flat 4D EGB black hole}
	
	The static spherically symmetric solution with the vanishing cosmological constant in (\ref{eqg00}) has been derived  \cite{GlLi20} as the form
	\begin{equation}\label{eqg00wL}
		g_{00}(r)=1+\frac{r^2}{2\alpha} \Big[1- \sqrt{1+\frac{8 \alpha M}{r^3}}\Big],
	\end{equation}
	\\ 
	where $M$ is the positive gravitational mass. At a large distance, the 4D EGB metric reduces to the Schwarzschild solution as follows
	\begin{equation}
		g_{00}(r)\simeq1-\frac{2 M}{r}.
	\end{equation}
	 \\
	 In Ref. \cite{GlLi20}, the authors argued that the GB term solves the problem of singularity by calculating Ricci scalar and arguing that the gravitational forces at short distances are repulsive and the particle cannot reach a $r=0$ point. To insure, one can calculate the Ricci scalar and the Kretschmann scalar and show whether the GB term alone removes the singularity or not.\\
	According to Eq. (\ref{eqg00wL}), there are two horizons located at
	\begin{equation}
		r_{\pm}=M\Big[1\pm\sqrt{1-\frac{\alpha}{M^2}}\Big],
	\end{equation}
	\\
	$r_-$ is the horizon of the white hole which has been covered with the horizon of a black hole, $r_+$. Hence, there is the critical mass that dependencies to $\alpha$ as follows
	
	\begin{equation}\label{Mcri}
		M_{cri}=\sqrt{\alpha}.
	\end{equation}
	
	When the gravitational mass of a black hole is greater than a square root of the Gauss-Bonnet coupling constant ($M>M_{cri}$), the black hole has two horizons: the white hole horizon, the black hole horizon, and the gravitational potential has a minimum located at $r_0$. Then, reducing mass to the critical mass ($M=M_{cri}$), white hole and black hole horizons would be merged together. If $M<M_{cri}$, there is no horizon but the minimum exists yet. Anyway, the important prediction of 4D EGB spacetime is the stopping collapse when the size of the black hole reaches $r_0$, where the attraction of the black hole is balanced to repulsive gravitational force. We demonstrate $g_{00}$ versus $r$ in three situations and compare these with the prediction of general relativity.\\
	We will calculate the temperature of the EGB black holes with tunneling from the horizon of the black hole in section $III$.\\

	\begin{figure}[ht]
		\centering
		\includegraphics [height=7 cm] {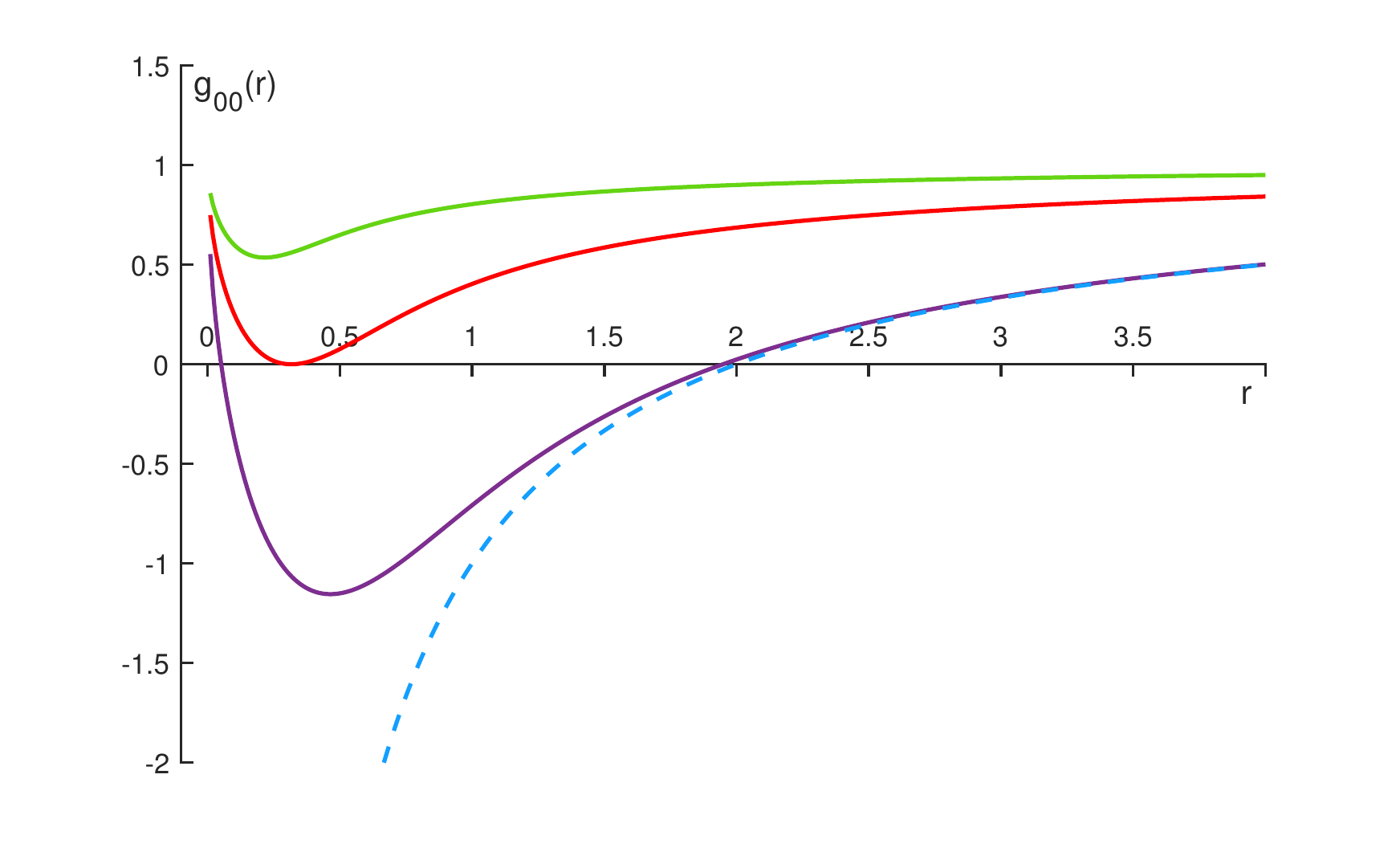}
		\caption{\scriptsize{Plot of $g_{00}(r)$ versus $r$ for asymptotically flat EGB black hole (with $\alpha=0.1$). Purple line: $M>M_{cri}$, there is the white hole and the black hole horizons. Red line: $M=M_{cri}$, there is the merged horizon. Green line: $M<M_{cri}$, there is no horizon. In three situation, there are a minimum in gravitatinal potentioal located at $r_0$ where the attractive and repulsive gravitational force is balanced. Dashed line shows general relativity pridiction.}}
		\label{figure_1}
	\end{figure}
	
	\subsection{Asymptotically de Sitter 4D EGB black hole}
	If we take the $\Lambda>0$ in the Eq. (\ref{eqg00}), we have the metric of the four dimensions of EGB as follows 
	
	\begin{equation}\label{g00dsEGB}
	g_{00}(r)=1+\frac{r^{2}}{2\alpha} \Big[1-\sqrt{1+\frac{8 M \alpha}{r^{3}}+\frac{4 \alpha \Lambda}{3}}\Big].
	\end{equation}
	
\begin{figure}[ht]
		\centering
		\includegraphics[height=3 cm,width=3.7 cm]  {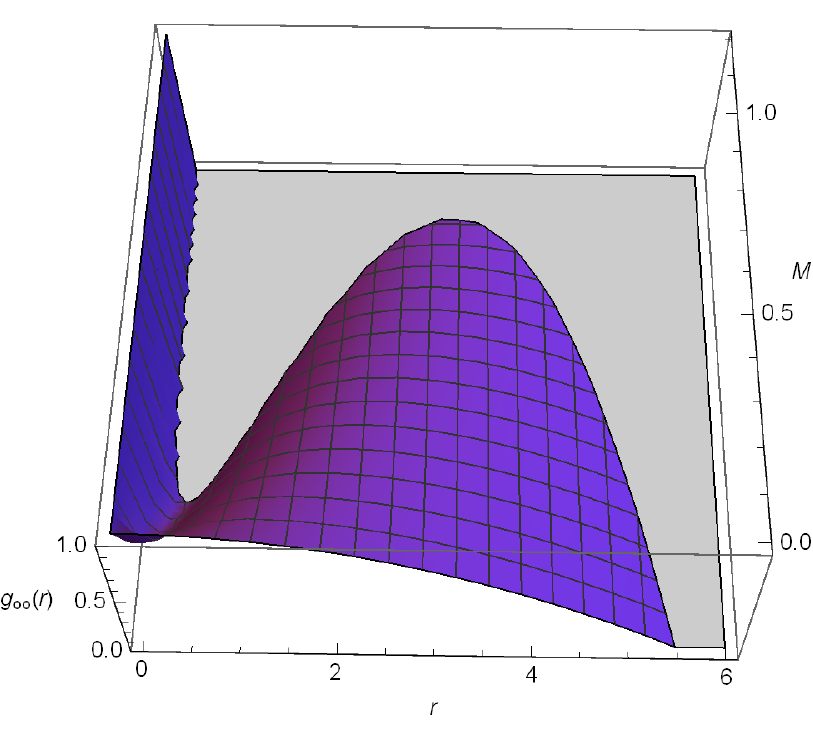}
		\hspace*{0.1cm}
		\includegraphics[height=3 cm,width=3.7 cm] {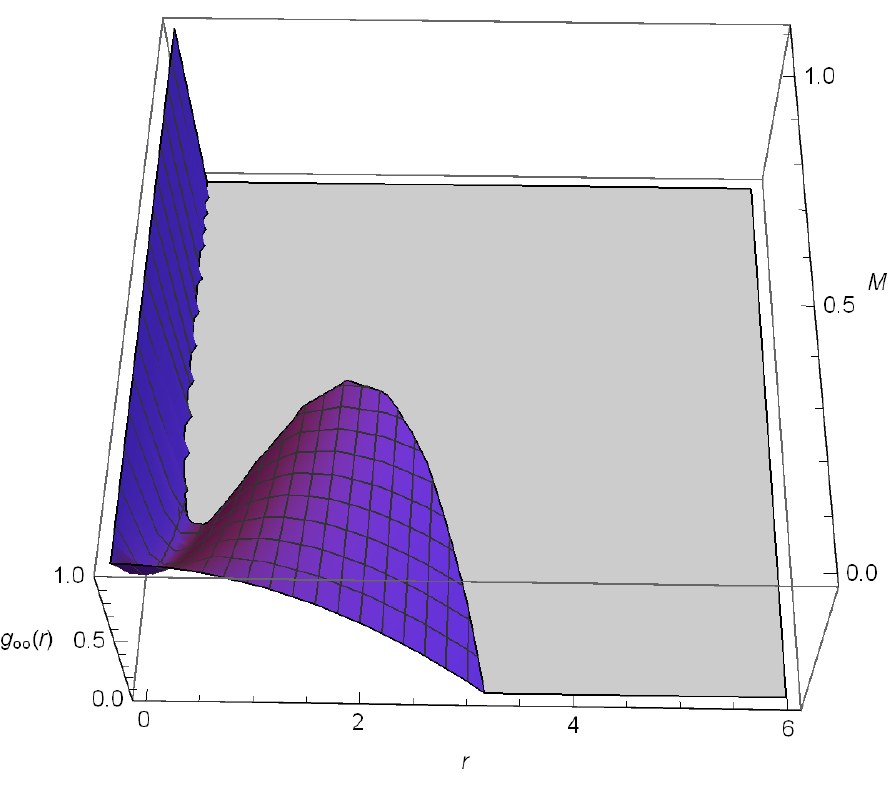}
		\hspace*{0.12cm}
		\includegraphics[height=3 cm,width=3.7cm]  {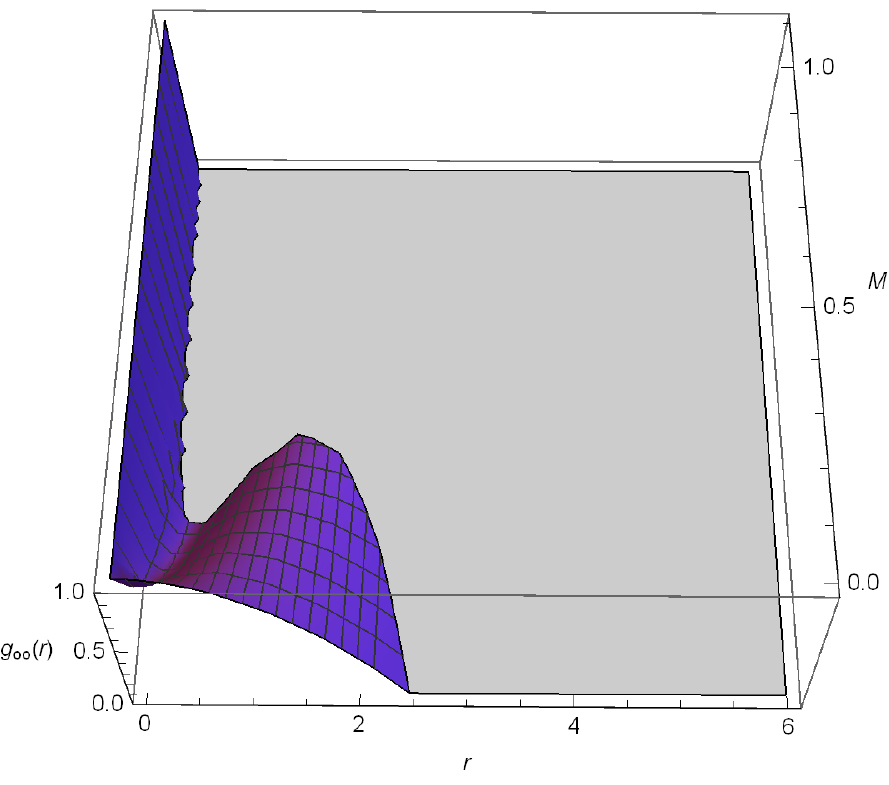}
		\hspace*{0.12cm}
		\includegraphics[height=3 cm,width=3.7cm]  {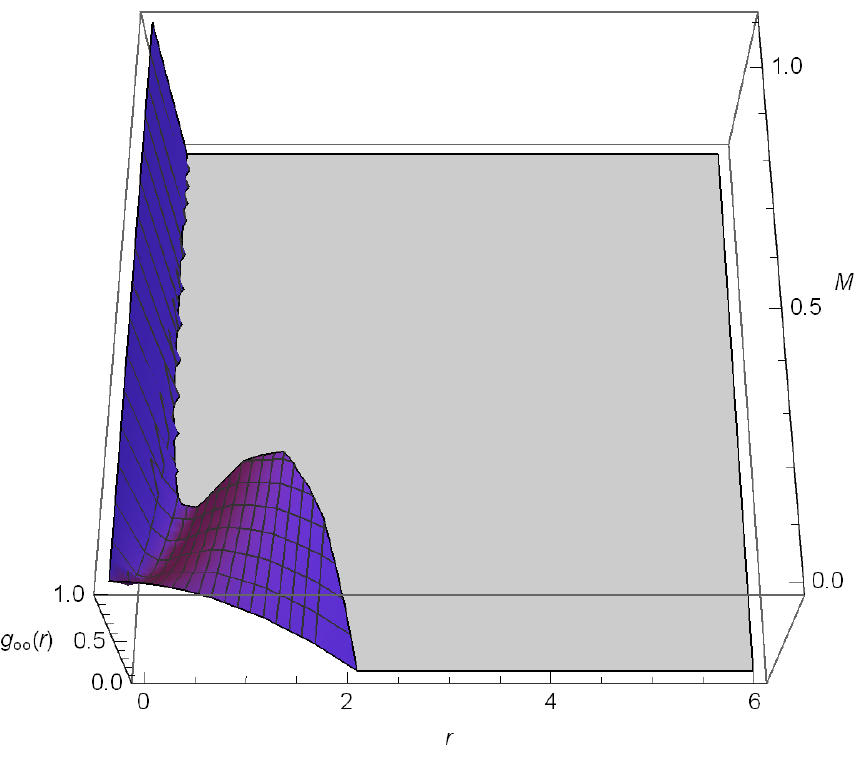}
		\vspace*{0.5cm}
		\includegraphics[height=3 cm,width=3.7 cm]  {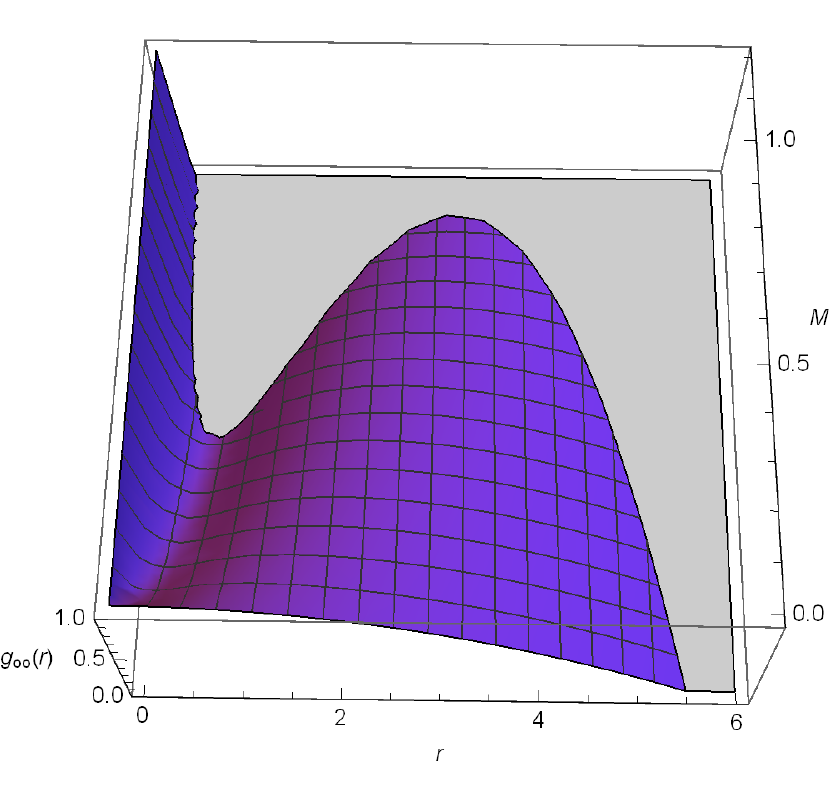}
		\hspace*{0.12cm}
		\includegraphics[height=3 cm,width=3.7 cm] {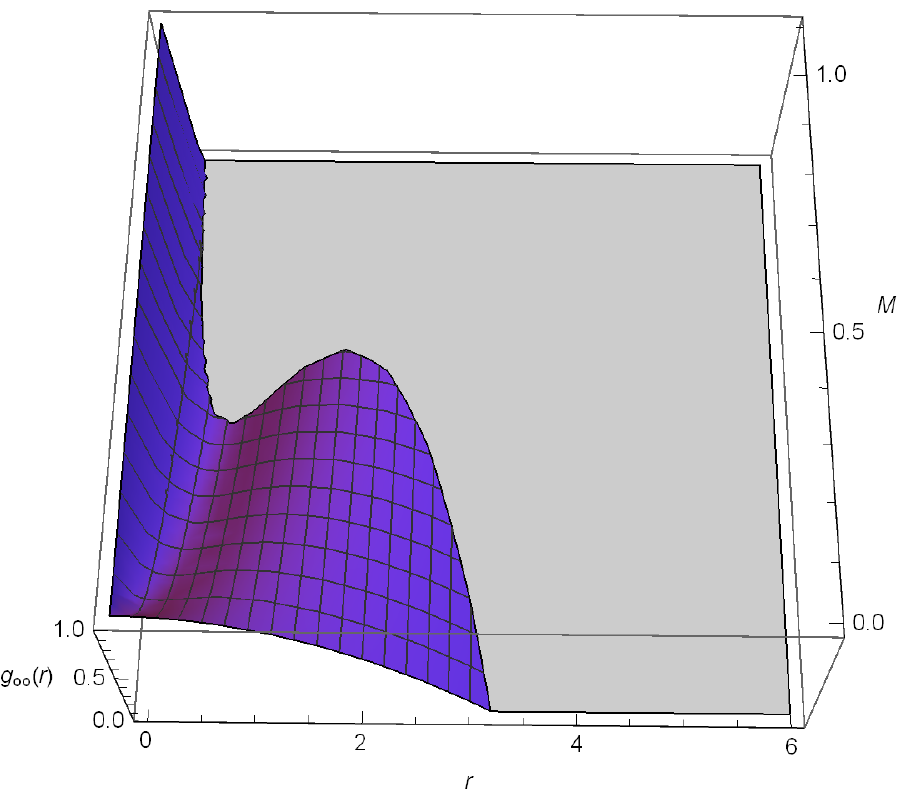}
		\hspace*{0.12cm}
		\includegraphics[height=3 cm,width=3.7cm]  {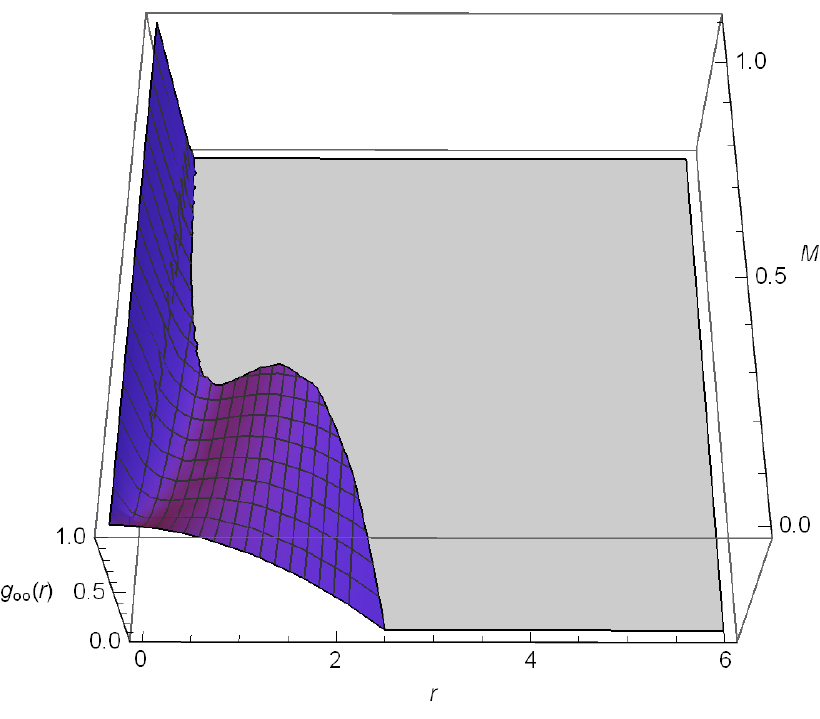}
		\hspace*{0.12cm}
		\includegraphics[height=3 cm,width=3.7cm]  {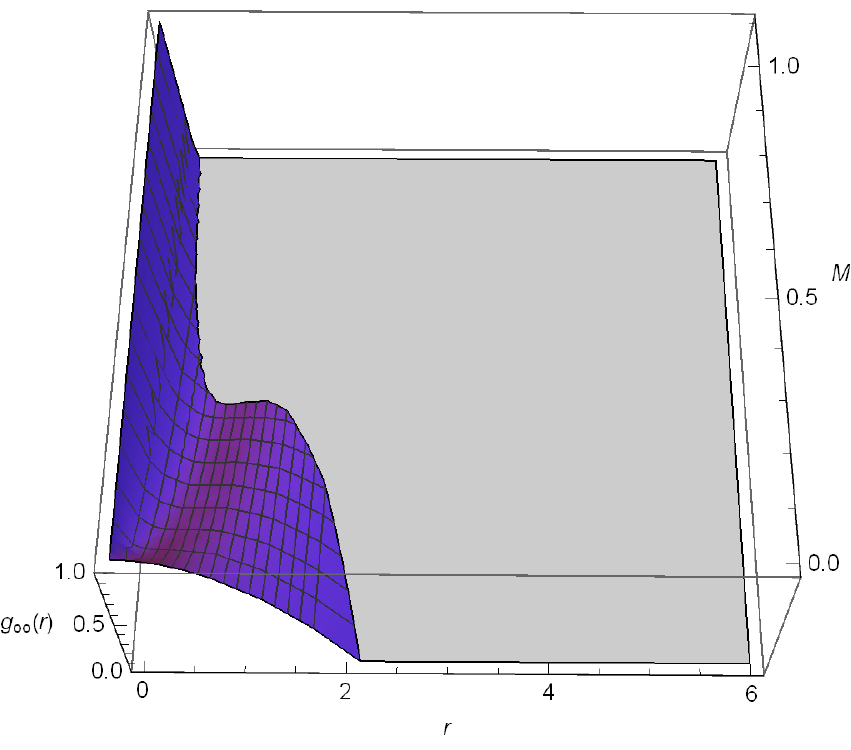}
		\vspace*{0.5cm}
		\includegraphics[height=3 cm,width=3.7 cm]  {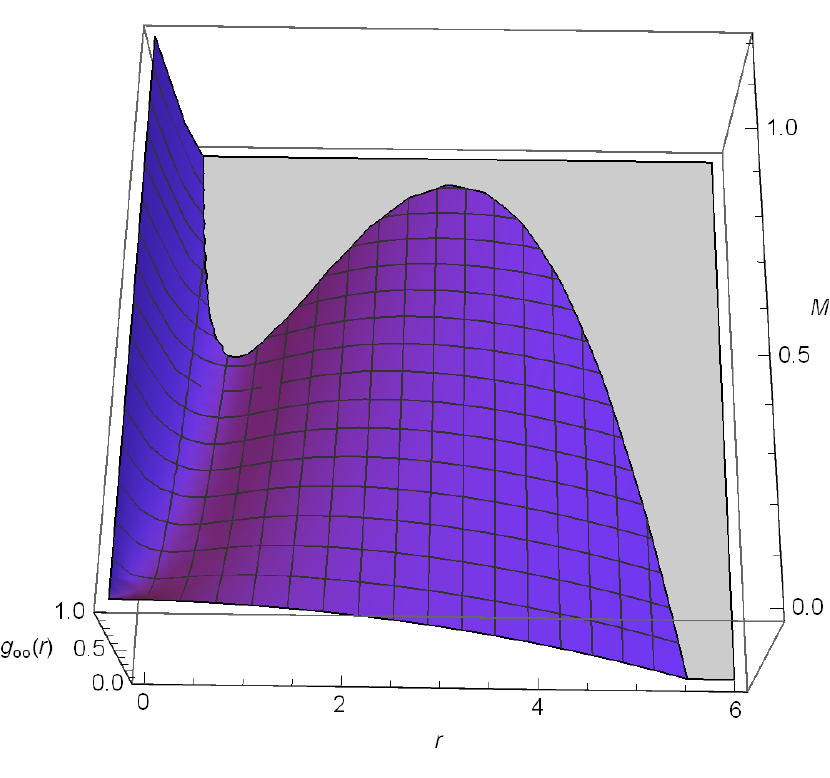}
		\hspace*{0.12cm}
		\includegraphics[height=3 cm,width=3.7 cm] {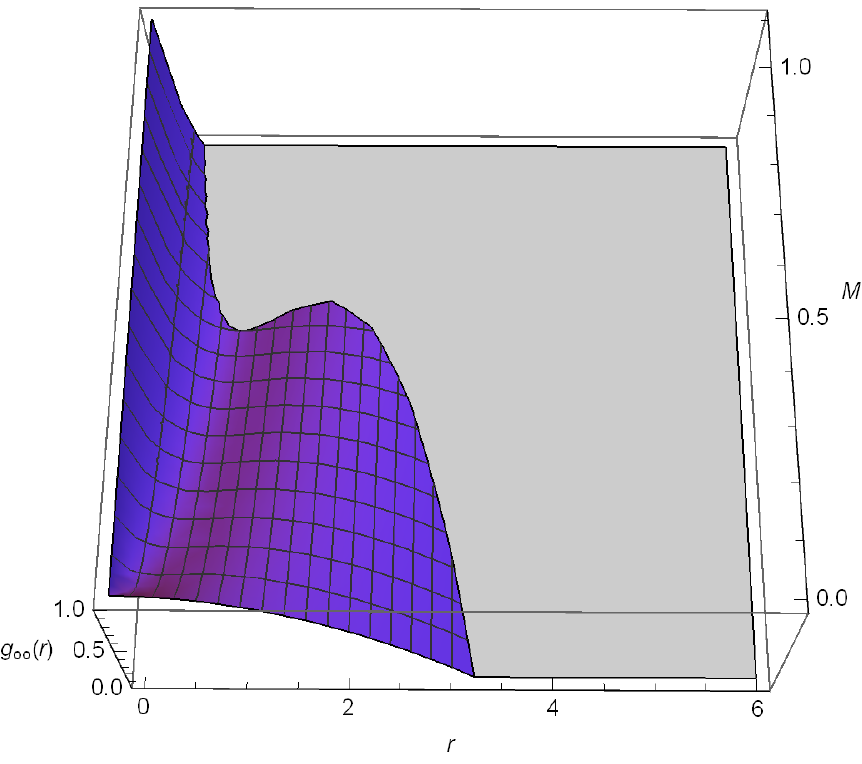}
		\hspace*{0.12cm}
		\includegraphics[height=3 cm,width=3.7cm]  {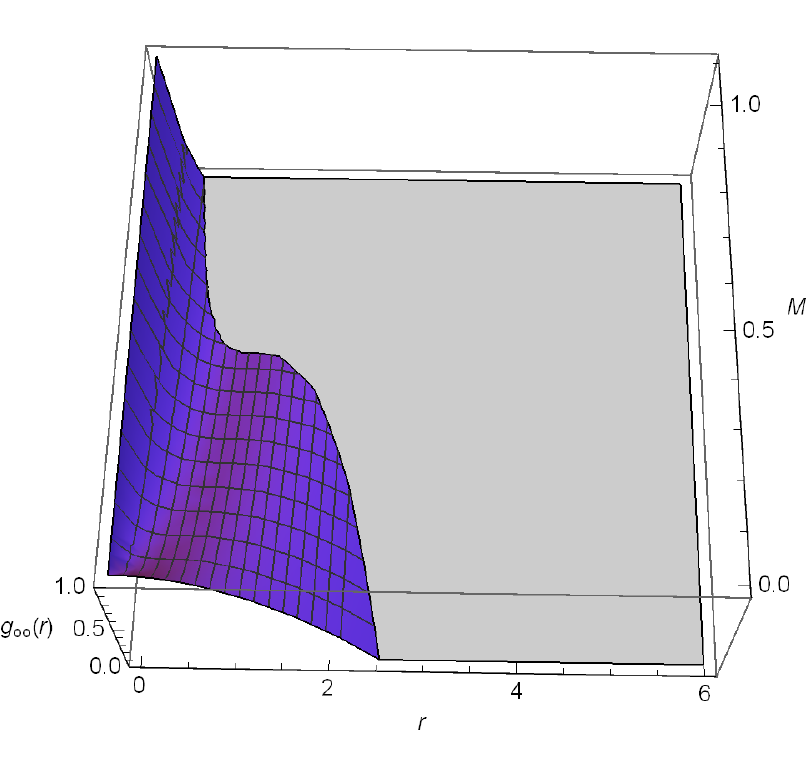}
		\hspace*{0.12cm}
		\includegraphics[height=3 cm,width=3.7cm]  {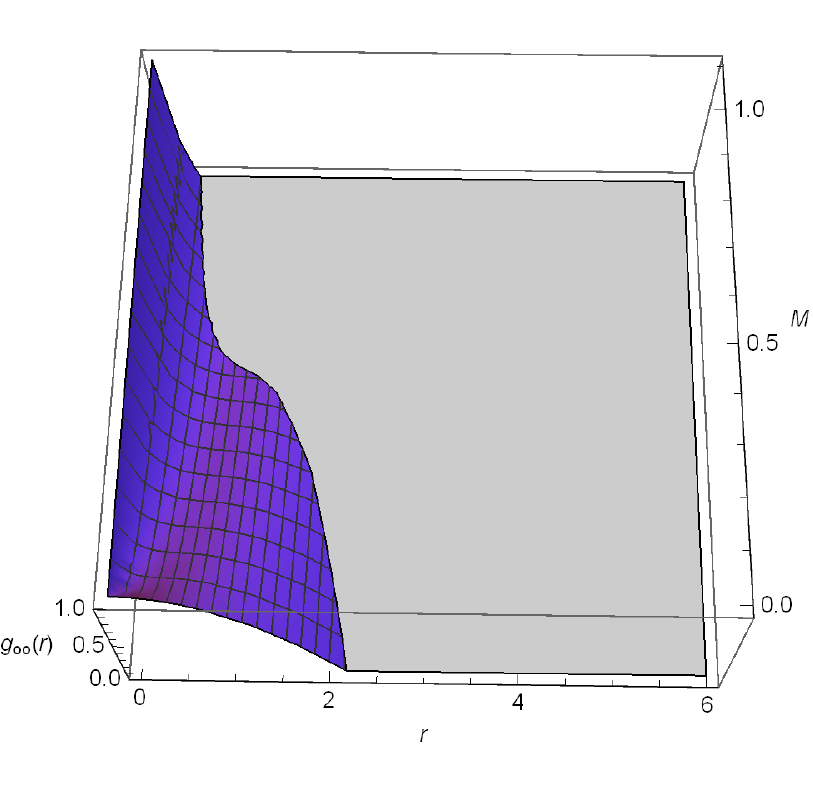}
		\vspace*{0.5cm}
		\includegraphics[height=3 cm,width=3.7 cm]  {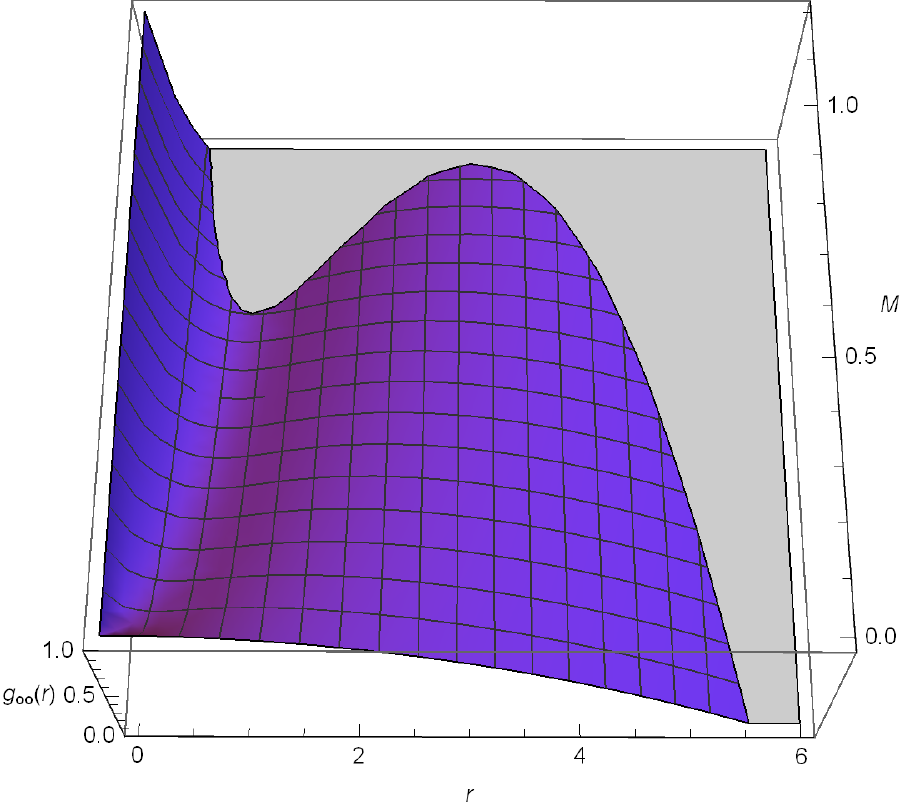}
		\hspace*{0.12cm}
		\includegraphics[height=3 cm,width=3.7 cm] {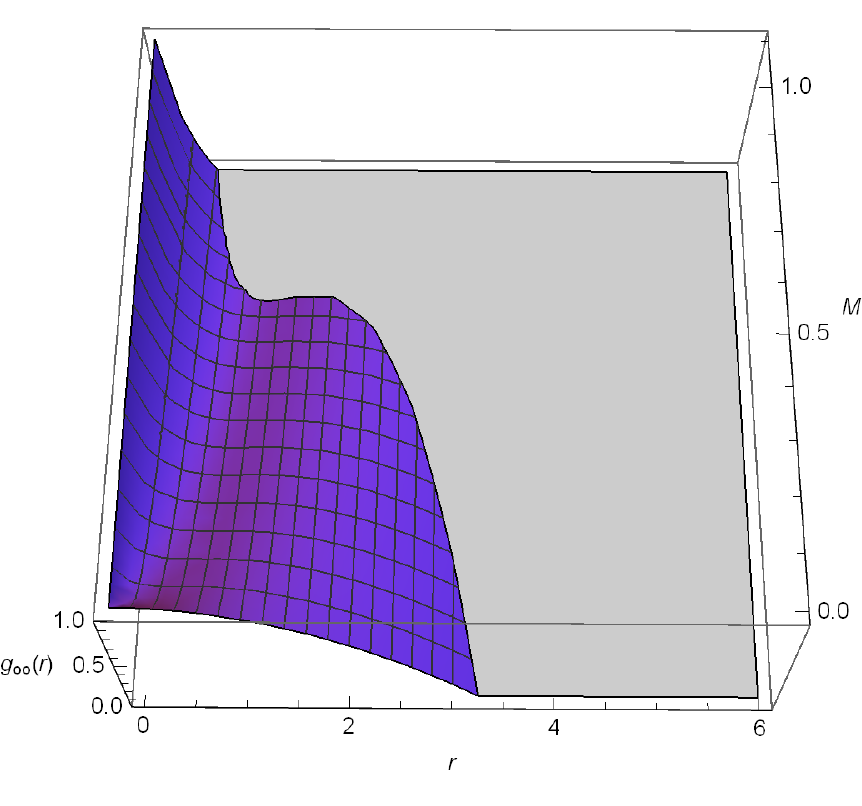}
		\hspace*{0.12cm}
		\includegraphics[height=3 cm,width=3.7cm]  {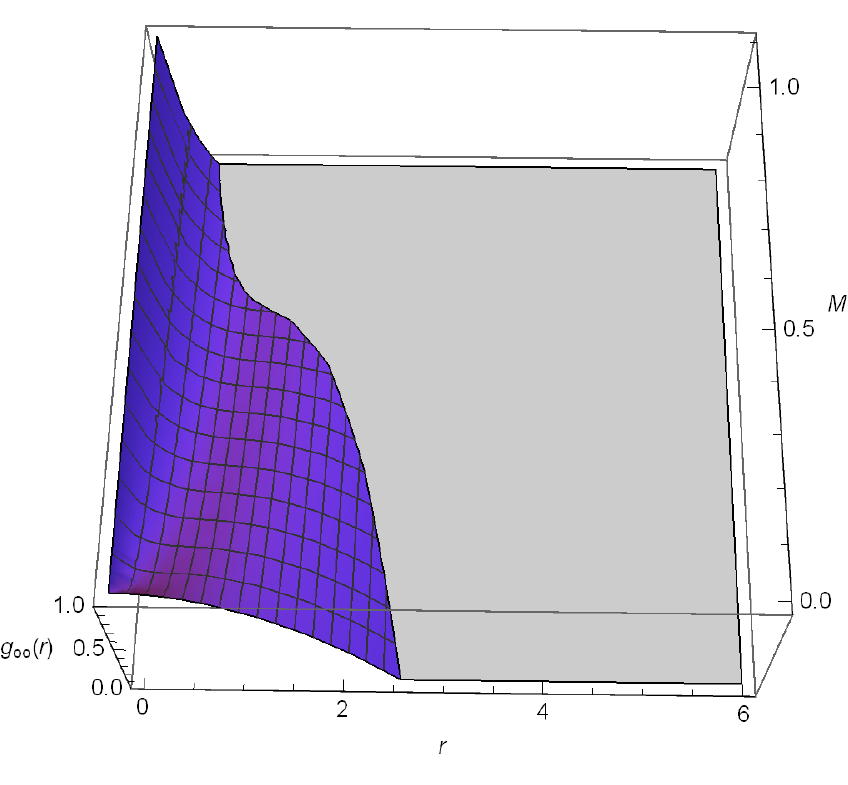}
		\hspace*{0.12cm}
		\includegraphics[height=3 cm,width=3.7cm]  {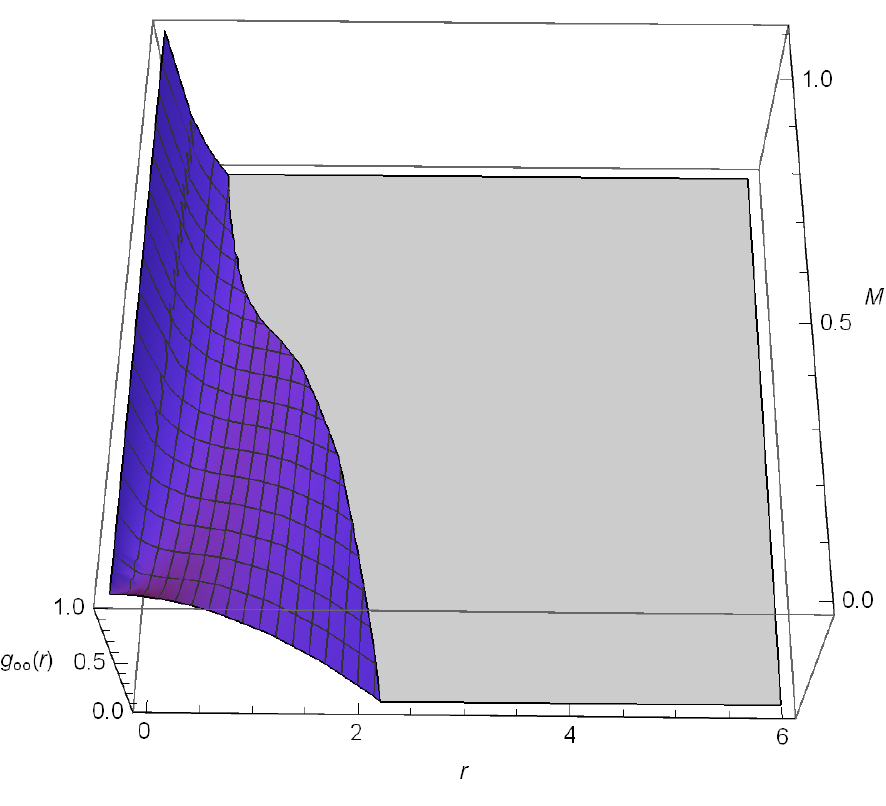}
		\vspace*{0.5cm}
		\caption{\scriptsize{ $g_{00}(r)$ versus of $M$ and $r$ with different $\alpha$ and $\Lambda$ for asymptotically dS EGB black hole. Plots have been depicted with $\alpha=0.1,0.3,0.5,0.7$ from up to down and $\Lambda=0.1,0.3,0.5,0.7$ from left to right. As regards each row, as $\Lambda$ increases, the $M_{cri2}$ decreases and the mass range in which there are three horizons becomes more limited. Also, as regards each colum, as $\alpha$ increases, the $M_{cri1}$ increases too, and the mass range in which there are three horizons becomes more limited.}}
		\label{figure_2}
	\end{figure}

	To study tunneling, we first need to know the structure of spacetime and the horizons of black holes. We need to know exactly where the horizon is located and what the barrier is the particle tunnels through. To do this, we must examine the horizons of this metric. As regard as Eq. (\ref{g00dsEGB}), radius of horizon determines with three-parameter $M,\alpha$ and $\Lambda$. To describe effects of the various $M,\alpha$ and $\Lambda$, we illustrate $g_{00}(r)$ versus $M$ and $r$ with different $\alpha$ and $\Lambda$ in Fig. (\ref{figure_2}). Figure has been depicted with the $\alpha=0.1,0.3,0.5,0.7$ from up to down and $\Lambda=0.1,0.3,0.5,0.7$ from left to right. Considering the first diagram (top and left one) which has been depicted with $\alpha=0.1$ and $\Lambda=0.1$, there is two critical mass, when $M=M_{cri1}$, the inner horizon and the black hole horizon merge and when $M=M_{cri2}$, the black hole horizon and the cosmological horizon merge. As a result, for a finite range of the mass, there are three horizons: the inner horizon ($r_{-}$), the black hole horizon ($r_{BH}$), and the cosmological horizon ($r_{CH}$) such that $r_{-}<r_{BH}<r_{CH}$ (see Fig. 3). Cases with $M<M_{cri1}$ or $M>M_{cri2}$ are not the physical cases with conventional definitions; one can be considered as a star without a black hole horizon and another one can be considered as a mass density distributes to a cosmological scales that is inconsistent with the assumption of solving vacuum equations. It is necessary to emphasize that we will take the range of the mass that three horizons exist. On the other hand, to check the effect of $\Lambda$, as regards each row in Fig. 2, as $\Lambda$ increases, the $M_{cri2}$ decreases and the mass range in which there are three horizons becomes more limited. Also, as regards each column in Fig. 2, as $\alpha$ increases, the $M_{cri1}$ increases too, and the mass range in which there are three horizons becomes more limited. In the following, in section III we will probe into the tunneling process from the black hole and the cosmological horizons.\\

	\begin{figure}[ht]
		\centering
		\includegraphics [height=7 cm] {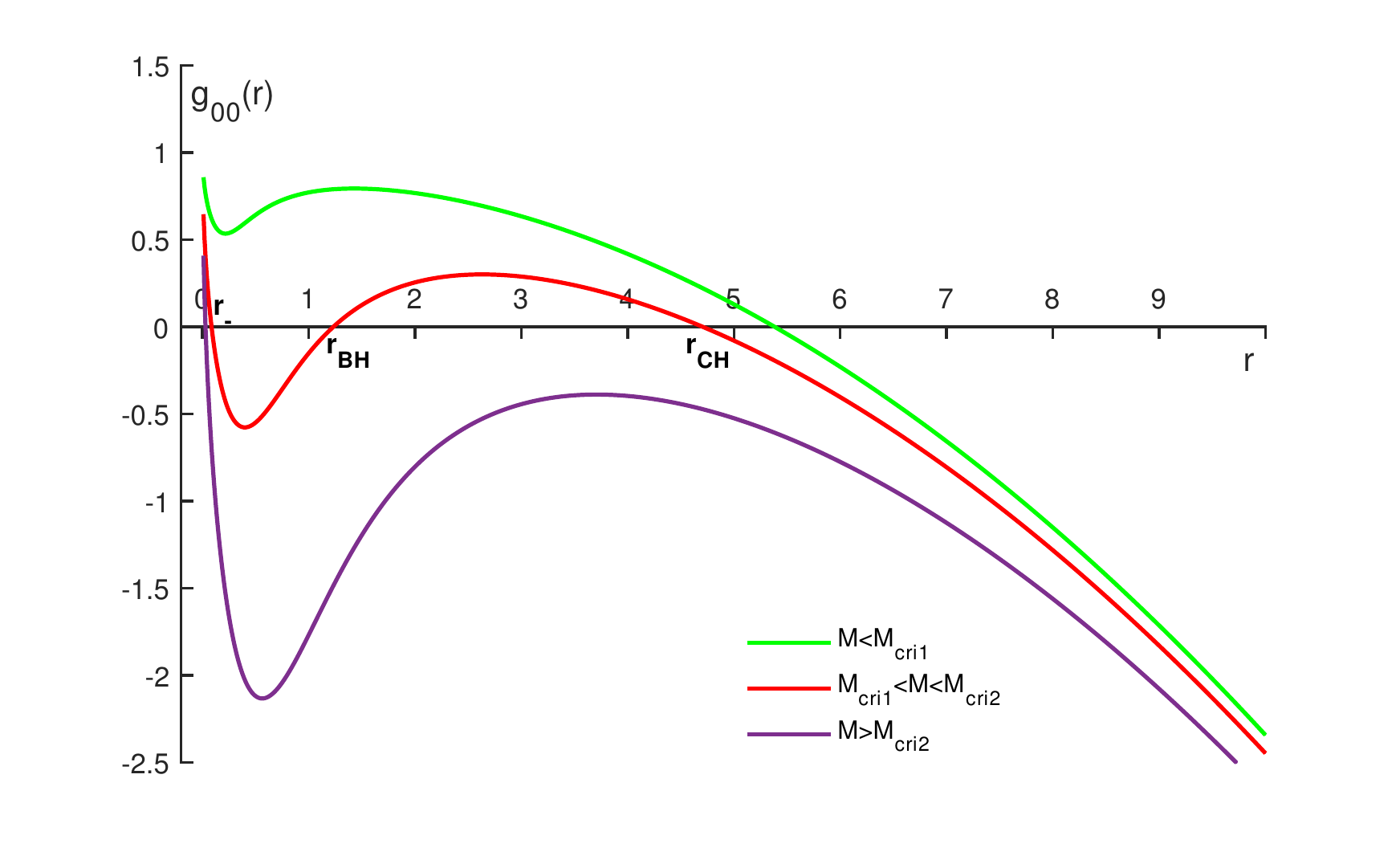}
		\caption{\scriptsize{Plot of $g_{00}(r)$ versus $r$ for the asymptotically ds EGB black hole. Purple line: $M>M_{cri2}$, Red line: $M_{cri1}<M<M_{cri2}$, Green line: $M<M_{cri1}$. Plot depicted with $\alpha=0.1$ and $\Lambda=0.1$. For a finite range of the mass, there are three horizons: the inner horizon ($r_{-}$), the black hole horizon ($r_{BH}$) and the cosmological horizon ($r_{CH}$)}}
		\label{figure_3}
	\end{figure}

	\subsection{Asymptotically Anti-de Sitter 4D EGB black hole}
	Drawing the spacetime with the negative cosmological constant in 4D EGB gravity is the issue which we will answer in this subsection. So far, a lot of research has been done on 5D AdS EGB black holes \cite{Cai02,Noj02,Kum19,AroEs19} and charged AdS 4D EGB black holes, too \cite{PFer20,EsJa20,GhSin21,JafZan09,Hed20}. However, there is a gap in a comprehensive study of asymptotically AdS 4D EGB black holes and a comparison with asymptotically flat and dS 4D EGB black holes.\\
	To construct the AdS 4D EGB black hole, we take $\Lambda<0$ in Eq. \ref{eqg00}. As a result, $g_{00}(r)$ becoms as follows
	
	\begin{equation}\label{eqg00Ads}
		g_{00}(r)=1+\frac{r^{2}}{2\alpha} \Big[1-\sqrt{1+\frac{8 M \alpha}{r^{3}}-\frac{4 \alpha \Lambda}{3}}\Big].
	\end{equation}
	
\begin{figure}[ht]
		\centering
		\includegraphics [height=7 cm] {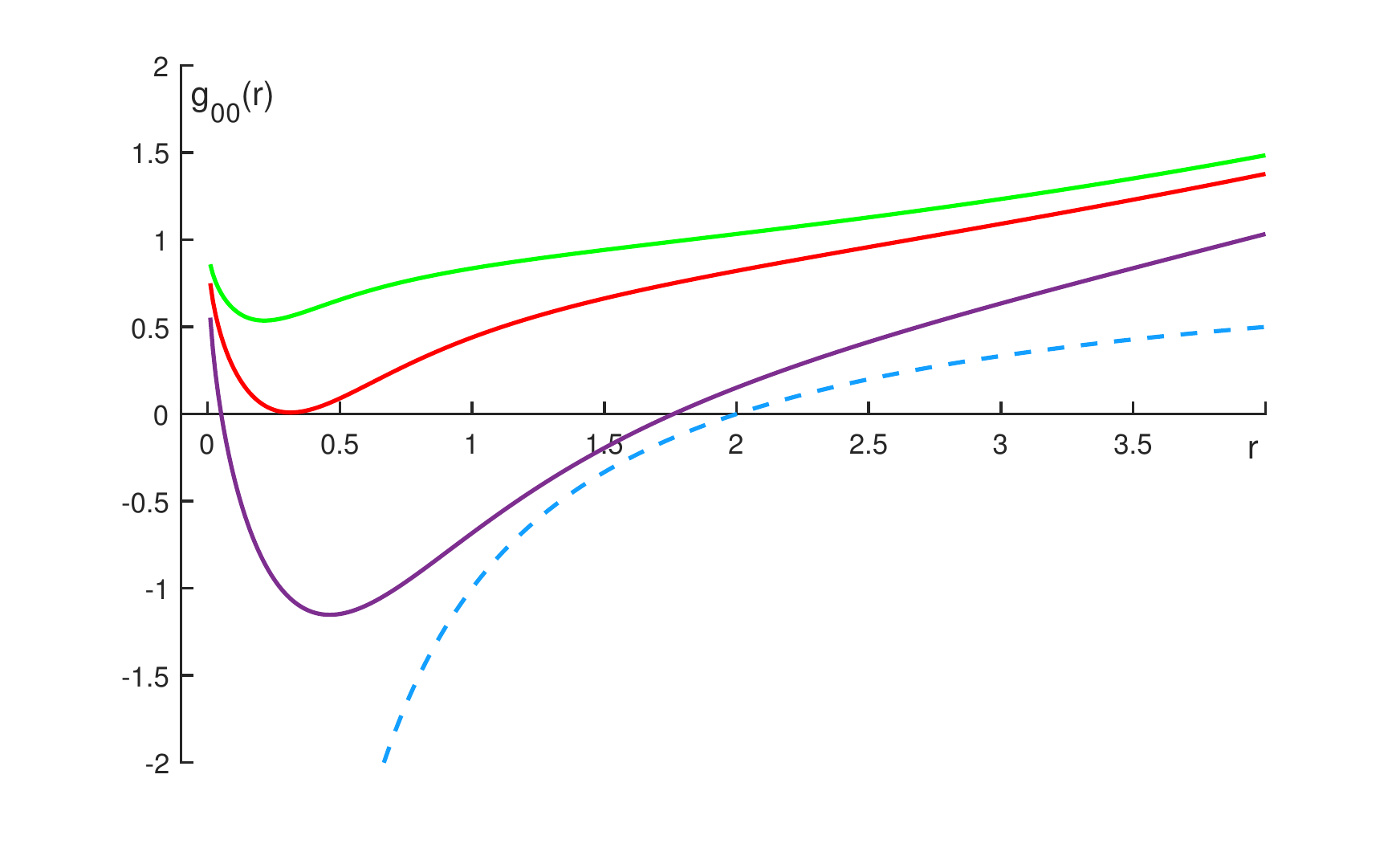}
		\caption{\scriptsize{Plot of $g_{00}(r)$ versus $r$ for the AdS 4D EGB black hole. AdS 4D EGB pridiction has been depicted from up to down with $M<M_{cri}$ (Green), $M=M_{cri}$ (Red), $M>M_{cri}$ (Purpule). Plot has been depicted with $\alpha=0.1$ and $\Lambda=0.1$. Dashed line shows GR pridiction.}}
		\label{figure_4}
	\end{figure}

\begin{figure}[ht]
		\centering
		\includegraphics[height=5 cm,width=7 cm]  {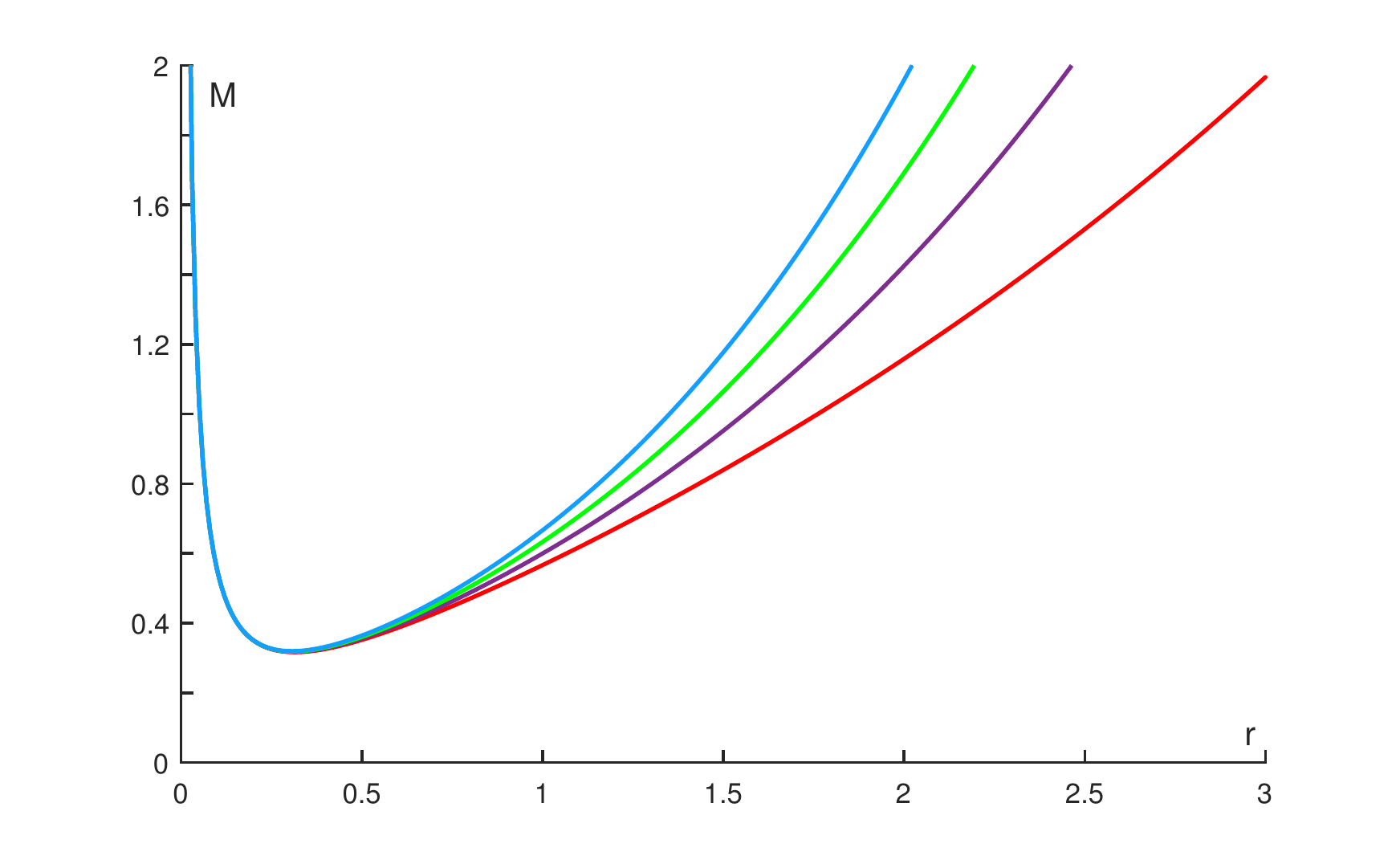}
		\hspace*{1cm}
		\includegraphics[height=5 cm,width=7 cm] {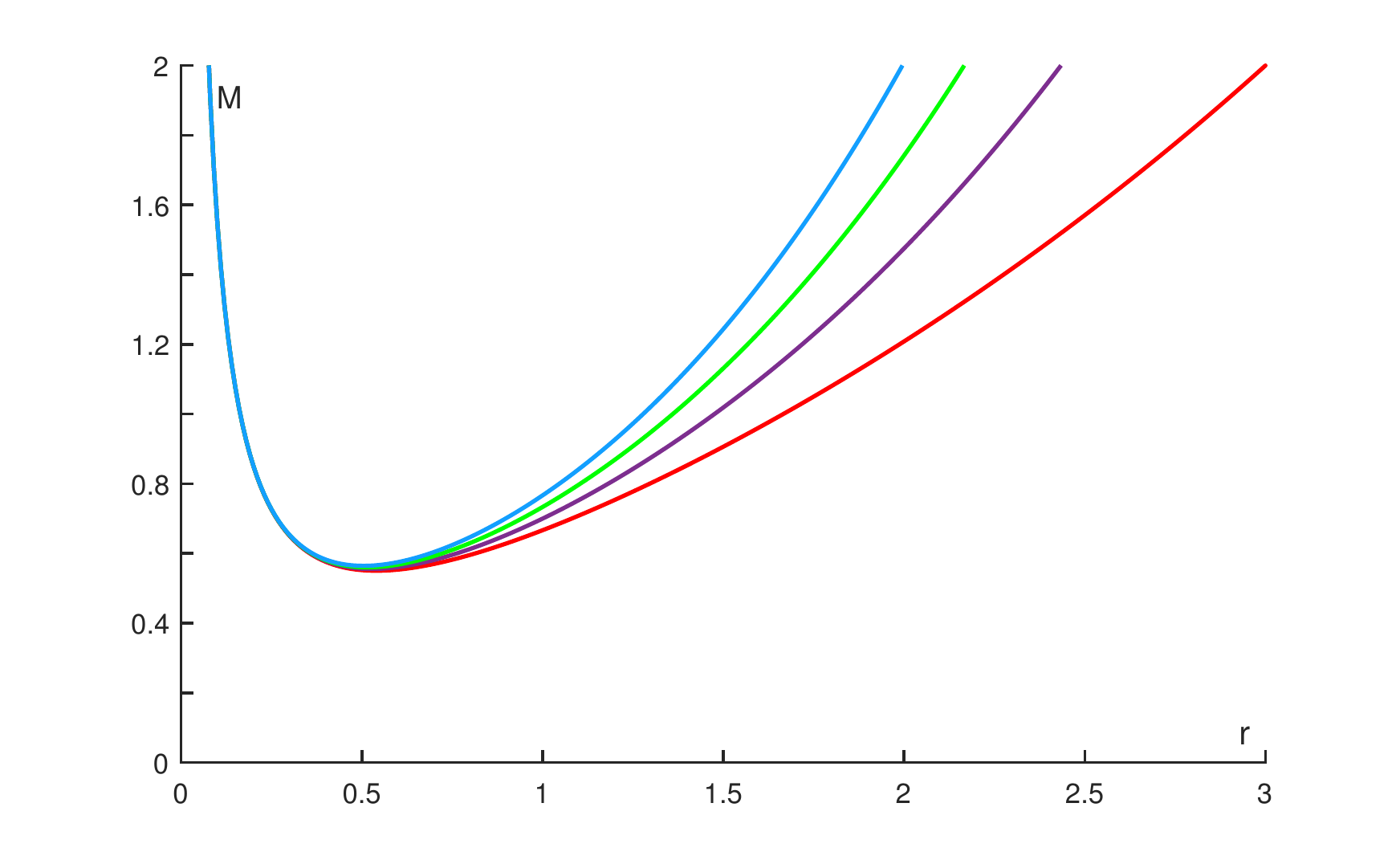}
		\caption{\scriptsize{The behavior of mass versus radius for the AdS 4D EGB black hole. Left plot has been depicted with $\alpha=0.1$ and right one with $\alpha=0.3$ while $\Lambda=0.7, 0.5, 0.3 0.1$ from up to down. These plots illustrate that $\Lambda$ changing affects the asymptotic behavior.}}
		\label{figure_5}
	\end{figure}

\begin{figure}[ht]
		\centering 
		\includegraphics[height=5 cm,width=7 cm]  {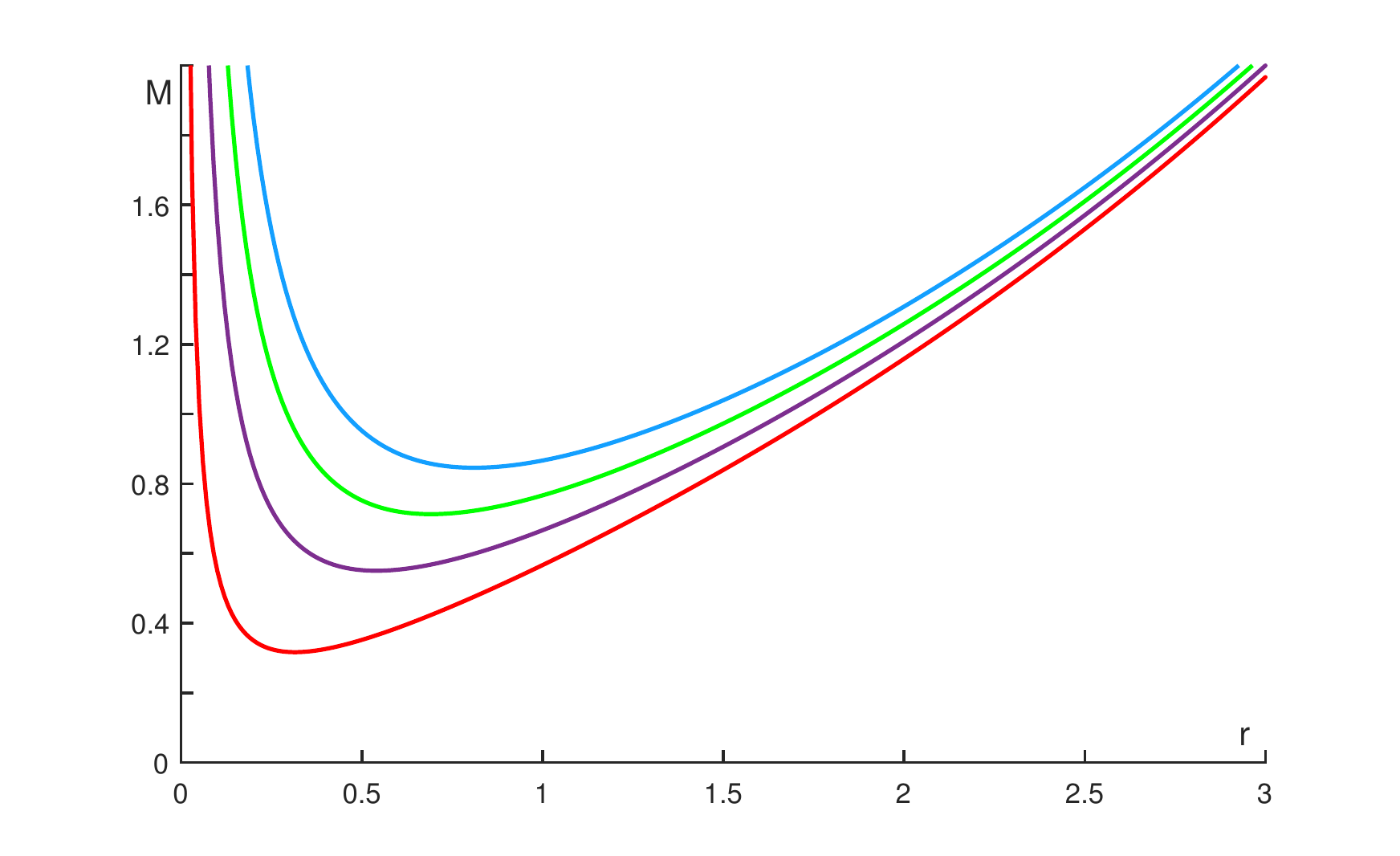}
		\hspace*{1cm}
		\includegraphics[height=5 cm,width=7 cm] {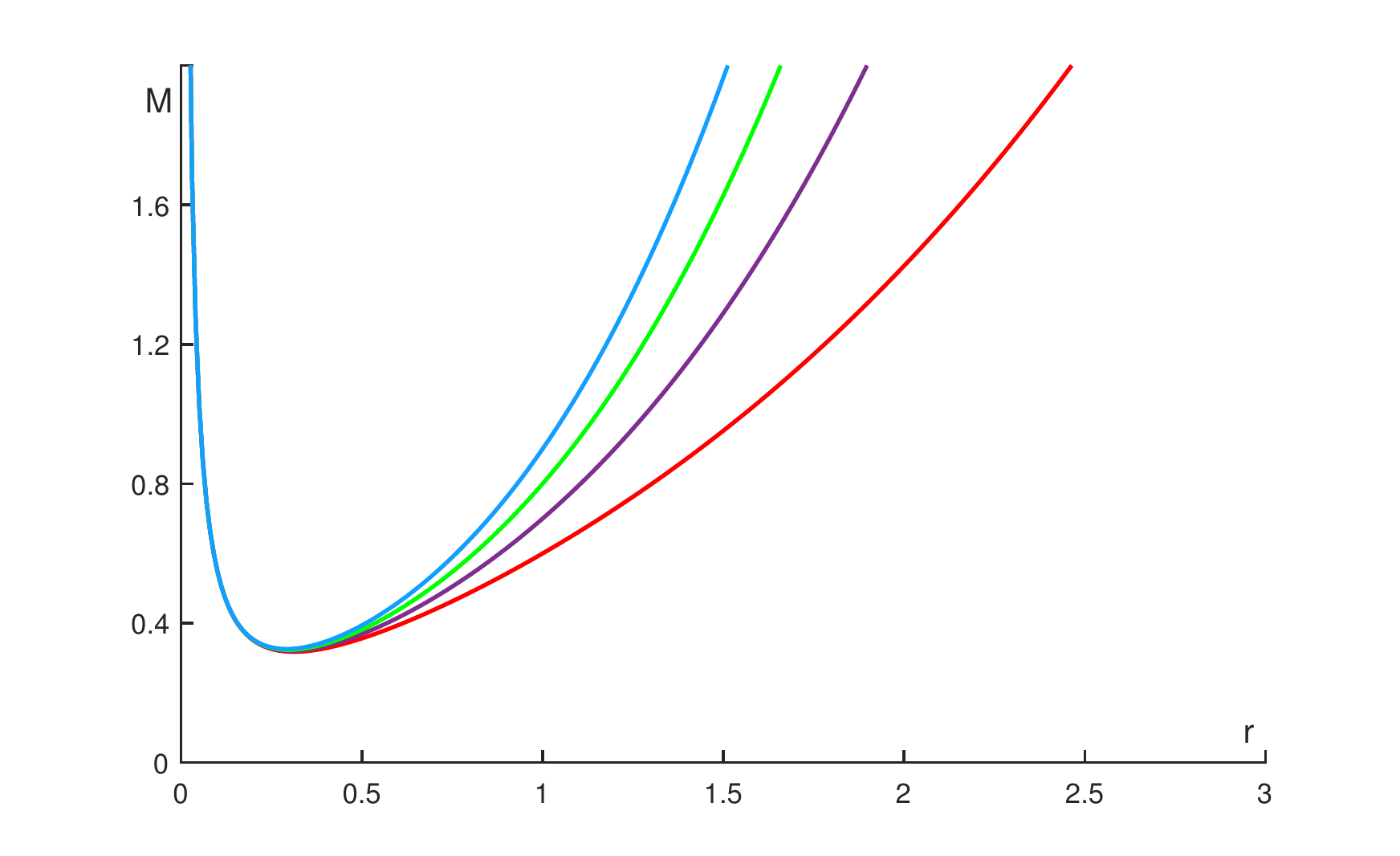}
		\vspace*{0.5cm}
		\caption{\scriptsize{The behavior of mass versus radius of AdS 4D EGB black hole. Left plot has been depicted with $\Lambda=0.1$ and right one with $\Lambda=0.3$ while $\alpha=0.7, 0.5, 0.3 0.1$ from up to down. Decreasing $\alpha$ decreases the minimum mass.}}
		\label{figure_6}
	\end{figure}

	To mapping the structure of spacetime and finding the effects of $M$, $\alpha$ and $\Lambda$, we plot Fig. \ref{figure_4}, Fig. \ref{figure_5} and Fig. \ref{figure_6}. In Fig. \ref{figure_5}, we fix $\alpha$ and $\Lambda$ and show that there is the critical mass (as an asymptotically flat). For $M>M_{cri}$, the black hole has two horizons: the black hole horizon and the cosmological horizon. In Fig. \ref{figure_5} and \ref{figure_6}, we illustrate the behavior of mass versus the radius of horizons with various $\alpha$ and $\Lambda$. Fig. \ref{figure_5} has been depicted with fixed $\alpha$ and various $\Lambda$ and Fig. \ref{figure_6} has been depicted with various $\alpha$ and fixed $\Lambda$. What can be seen from these diagrams is decreasing $\alpha$ decreases the minimum mass while increasing $\Lambda$ just affects asymptotic behavior, as expected. Compared with asymptotically flat 4D EGB black hole, even though AdS 4D EGB black hole predicts the less critical mass, but the general behavior of the metric in the small radii is the same as the flat one. For this reason, we do not examine the tunneling process from this black hole and predict that the temperature behavior of this black hole will be the same as the same of the asymptotically flat one.\\
	

	\section{Temperature Of The EGB Black Hole With Tunneling Process}
	In the intelligent approach that Parikh and Wiczeck have provided two issues are considerable: energy conservation and dynamical geometry \cite{Wil00,Par04,Kra95}. In this picture, particle, and antiparticle are created with zero total energy in the near of one side of the horizon, after that one particle tunnels from the horizon in a semiclassical way. A particle tunnels through a barrier created by the particle's energy itself. As a result, the radius and mass of the black hole reduce as much as the particle of energy $\omega$. The first step to the calculation of particle tunneling is the construction of the nonsingular line element on the horizon. To resolving horizon singularity, we use the Painlev\'{e} coordinate transformation with definition $t_p$ by $t_p=t-f(r)$. Applying this transformation on the metric (\ref{eqg00}), we will have a new nonsingular coordinate as follows
	\begin{equation}\label{ds2}
		ds^2=-g_{00}(r)dt_p^2+2 f'(r)g_{00}(r)dt_pdr+\big(\frac{1}{g_{00}(r)}-g_{00}(r)(f'(r))^2\big)dr^2+r^2d\Omega^2.
	\end{equation}
	Concerning to the constant-time slice should be flat, we impose the condition $\frac{1}{g_{00}(r)}-g_{00}(r)(f'(r))^2=1$.  Putting this restriction on the Eq.(\ref{ds2}), the metric (\ref{eqg00}) convert to
	\begin{equation}\label{ds2peinlev}
		ds^2=-g_{00}(r)dt_p^2+dr^2+2\sqrt{1-g_{00}(r)}\:dt_pdr+r^2d\Omega^2.
	\end{equation}
	Now, we present the primary calculation of the temperature of the black hole with the tunneling process. A particle is moving from an initial state in $r_{in}$ to the final state in $r_{out}$ as $r_{in}>r_{out}$. Tunneling calculation based on to account the imaginary part of the action for this particle as follows
	
	\begin{equation}\label{ImS000}
		\mathrm{Im}S\equiv \mathrm{Im}\int
		E\:dt=\mathrm{Im}\int_{r_{in}}^{r_{out}}
		p_{r}\:dr=\mathrm{Im}\int_{r_{in}}^{r_{out}}\int_0^{p_{r}}\:d\tilde{p_r}\:dr,
	\end{equation}
	
	where $r_{in}=r_{H}-\epsilon$ and $r_{out}= r'_{H}+\epsilon$, $\tilde{\omega}$ is particles' energy which is known as a self intraction. Putting Hamilton equation, $dp_r=\frac{dH}{\dot{r}}$, on the  Eq. (\ref{ImS000}), we have
	
	\begin{equation}\label{ImSgeneral}
		\mathrm{Im} S=\mathrm{Im}\int_{r_{in}}^{r_{out}}\int_{M}^{M-\omega}\frac{dH}{\dot{r}}\:dr=-
		\mathrm{Im} \int_0^\omega\int_{r_{in}}^{r_{out}}\frac{dr}{\dot{r}}\:d\tilde{\omega}.
	\end{equation}
	
	Because we note the massless particles' tunneling, we determine the light-like geodesics regarded to transformed metric (Eq. (\ref{ds2peinlev})) as follows
	
	\begin{equation}
		\dot{r}_p^2 +2\sqrt{1-g_{00}(r)}\:\dot{r}_p-g_{00}(r)=0.
	\end{equation}
	
	As a result, the light-like geodesics for crossed massless particles is given by
	
	\begin{equation}\label{rdot000}
		\dot{r}_p=\pm1-\sqrt{1-g_{00}(r)},
	\end{equation}
	
	where $+$ and $-$ signs indicate the outgoing and ingoing geodesics, respectively. With considering outgoing trajectories and substituting Eq. (\ref{rdot000}) in Eq. (\ref{ImSgeneral}), we find the imaginary part of the action as follows 
	
	\begin{equation}\label{Imaction}
		\mathrm{Im} S=-\mathrm{Im}\int_0^\omega\int_{r_{in}}^{r_{out}}\frac{dr\:d\tilde{\omega}}{1-\sqrt{1-g_{00}}}\,.
	\end{equation}\\
	
	\subsection{Tunneling from Asymptotically Flat 4D EGB black hole}
\begin{figure}[ht]
		\centering
		\includegraphics [height=5 cm] {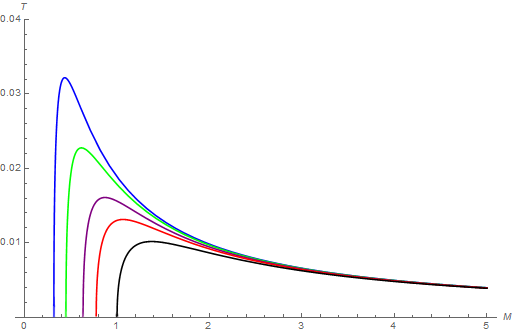}
		\caption{\scriptsize{Temperature of the asymptotically flat 4D EGB black hole versus Mass. The figure has been plotted with $\alpha=0.1, 0.2,0.4,0.6,1$ from up to down.}}
		\label{figure_7}
	\end{figure}

	The metric of EGB gravity which we examined in the previous section (Eq. (\ref{eqg00wL})), has two horizons located at
	\begin{equation}
		r_{BH}=M+\sqrt{M^{2}-\alpha},\qquad r_{WH}=M-\sqrt{M^{2}-\alpha}.
	\end{equation}
	In the tunneling process, particles created at the near of the inside of the black hole horizon located at $r_{in}=M+\sqrt{M^{2}-\alpha}-\epsilon$, then these tunnel from the black hole horizon and reach the outside of the black hole horizon located at $r_{out}=(M-\omega)+\sqrt{(M-\omega)^{2}-\alpha}+\epsilon$; where $\omega$ is the particles' energy and $\epsilon$ refers to the near amount of the horizon. To calculate the imaginary part of action, we replace Eq. (\ref{eqg00wL}) in Eq. (\ref{Imaction}), as follows
	\begin{equation}
		\mathrm{Im} S=-\mathrm{Im}\int_0^\omega\int_{r_{in}}^{r_{out}}\frac{dr\:d\tilde{\omega}}{1-\sqrt{\frac{r^2}{2\alpha} \Big[1\pm \sqrt{1+\frac{8 \alpha M}{r^3}}\Big]}}\,.
	\end{equation}
	This integral has two poles in the range of its limits. So, to deduce the poles we expand the denominator in terms of $r_{out}$ as follows
	\begin{equation}\label{dotr}
		\dot{r}_p=1-\sqrt{1-[g_{00}(r_{out})+g'_{00}(r_{out})(r-r_{out})-...]},
	\end{equation}
	where a prime indicates derivative with respect to $r$. By doing this, the pole in the range of the integral limits reducs to $r\to M-\omega+\sqrt{M^2-2 M \omega+\omega^2-\alpha}$ and the calculable integral with residue calculation becomes as follows
	\begin{eqnarray}\label{ImSFlat}
		&\mathrm{Im}S=-\mathrm{Im}\int_0^\omega\int_{r_{in}}^{r_{out}}
		\Bigg[1-\Bigg(1-\Big(\frac{6 (M-\tilde{\omega})}{\Big[M-\tilde{\omega}+\sqrt{(M-\tilde{\omega})^{2}-\alpha}\Big]^{2}\sqrt{1+\frac{8 (M-\tilde{\omega}) \alpha}{\Big[M-\tilde{\omega}+\sqrt{(M-\tilde{\omega})^{2}-\alpha}\Big]^{3}}}}\nonumber\\
		&-\frac{\Big[M-\tilde{\omega}+\sqrt{(M-\tilde{\omega})^{2}-\alpha}\Big] \Big(1-\sqrt{1+\frac{8(M-\tilde{\omega})\alpha}{\Big[M-\tilde{\omega}+\sqrt{(M-\tilde{\omega})^{2}-\alpha}\Big]^{3}}}\Big)}{\alpha}\Bigg)\nonumber\\
		&\Bigg(r-(M-\tilde{\omega}+\sqrt{(M-\tilde{\omega})^{2}-\alpha})\Big)\Bigg)^{1/2}\Bigg]^{-1}dr\:d\tilde{\omega}.
	\end{eqnarray}
	In the following, we compute the first integral with residue calculus and expand the result to the second order of $\omega$. 
	Calculating the second integral, we derive the imaginary part of the action in terms of particles' energy and the black hole mass. According to the relation between the emission rate, the imaginary part of the action, and the Boltzmann factor
	\begin{equation}\label{Gamma}
		\Gamma\simeq e^{-2 \mathrm{Im}S}=e^{-\beta\omega},
	\end{equation}
	the temperature of the black hole can be calculated since the Hawking temperature is the inverse of the Boltzmann factor, $T=1/\beta$. Finally, we derive the temperature of the black hole horizon in terms of the black hole mass and the coupling constant. We plot the temperature of the black hole horizon of the 4D EGB black hole as a function of the mass of the black hole in Fig. (\ref{figure_7}) with $\alpha=0.1, 0.2,0.4,0.6,1$. By reducing the mass of the black hole, the temperature increases to a maximum temperature when the black hole mass reaches a special mass. After, reducing mass, the temperature reduces, too. Eventually, the rest mass -the critical mass that has been introduced in Eq. (\ref{Mcri})- with merged horizons and zero temperature remains. On the other hand, by reducing the coupling constant, maximum temperature of the black hole horizion increases. Such behaviors of temperature are exactly matched with the behaviors of temperature that can be obtained from the standard definition of the Hawking temperature. Also, such results can be compared with the results of the thermodynamics for the AdS 4D EGB charged black hole \cite{PFer20, GhSin21} and the thermodynamics for AdS Lovelock gravity \cite{Kum19}. In addition, it is good to mention a comparison of the coupling constant effects on Hawking temperature of 4D EGB black hole as can be seen in papers such as Refs. \cite{ZhLi20, HHX21, GhMa17}.\\
		Furthermore, the existence of the second and higher order of the mass in the temperature relation illustrates the deviation from the pure thermal radiation \cite{ArzMed05,FirMan12}. Actually, if we suppose that the particle with energy $\omega_1$ and another particle with energy $\omega_2$ tunnel through the horizon, the correlation function between them can be calculated (based on Ref. \cite{ArzMed05}) by the following equation 
		\begin{equation}\label{correlation}
			\chi{(\omega_1+\omega_2;\omega_1,\omega_2)}=\ln{[\Gamma(\omega_1+\omega_2)]}-\ln{[\Gamma(\omega_1)\Gamma(\omega_2])}.
		\end{equation}
	Substituting the result of Eq. (\ref{ImSFlat}) in Eq. (\ref{Gamma}) and calculating the correlation function between the two particles (Eq. (\ref{correlation})), it can be seen that the correlation function is not zero due to the existence of nonlinear term of the mass in Eq. (\ref{ImSFlat}). In other words, Eq. (\ref{correlation}) results in the existence of a correlation between radiated mods and back-reaction effects. For the reasons mentioned in Refs. \cite{Par04, MVag05, ArzMed05, NozMe08}, if there is a correlation between emitted modes, then radiation deviates from pure thermal, unitary preserves, and loss information problem gets. Indeed, as regards the result of the Eq. (\ref{correlation}), whenever one particles' energy $\omega_1$ is radiated from the surface of the 4D EGB black hole, the deflections are created that influence the second radiated particles' energy $\omega_2$. On the other hand, as regards $\Gamma\sim\frac{\rho_{in}}{\rho_{out}}=exp (\Delta S)$ ($\rho$ indicates the density of the states), the correlation between the radiated modes can be associated with unitary and resolving the information loss problem \cite{MVag05, ArzMed05}. There is another direct way to investigate the unitarity by calculating the density matrix of Hawking radiation as the same has been done in Ref. \cite{SaSta15} which can be a proposal for another work. Therefore, the existence of the coupling constant alone prevents the temperature divergency and the unitary preserves due to the correlation between the radiated modes.\\
	
	\subsection{Tunneling from Asymptotically de Sitter 4D EGB black hole}
\begin{figure}[ht]
		\centering
		\includegraphics[height=4 cm,width=5.5 cm]  {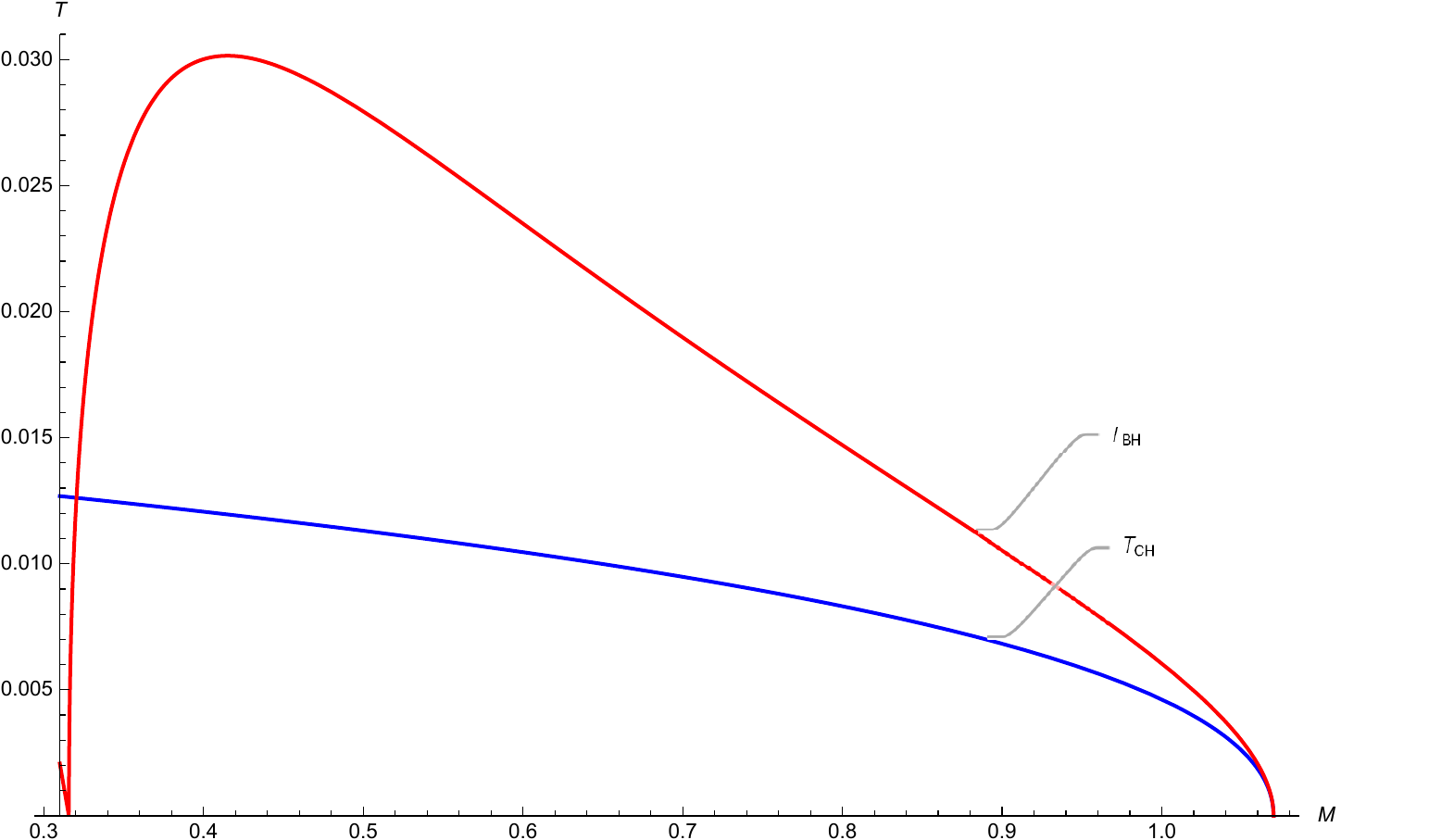}
		\hspace*{0.5cm}
		\includegraphics[height=4 cm,width=5.5 cm] {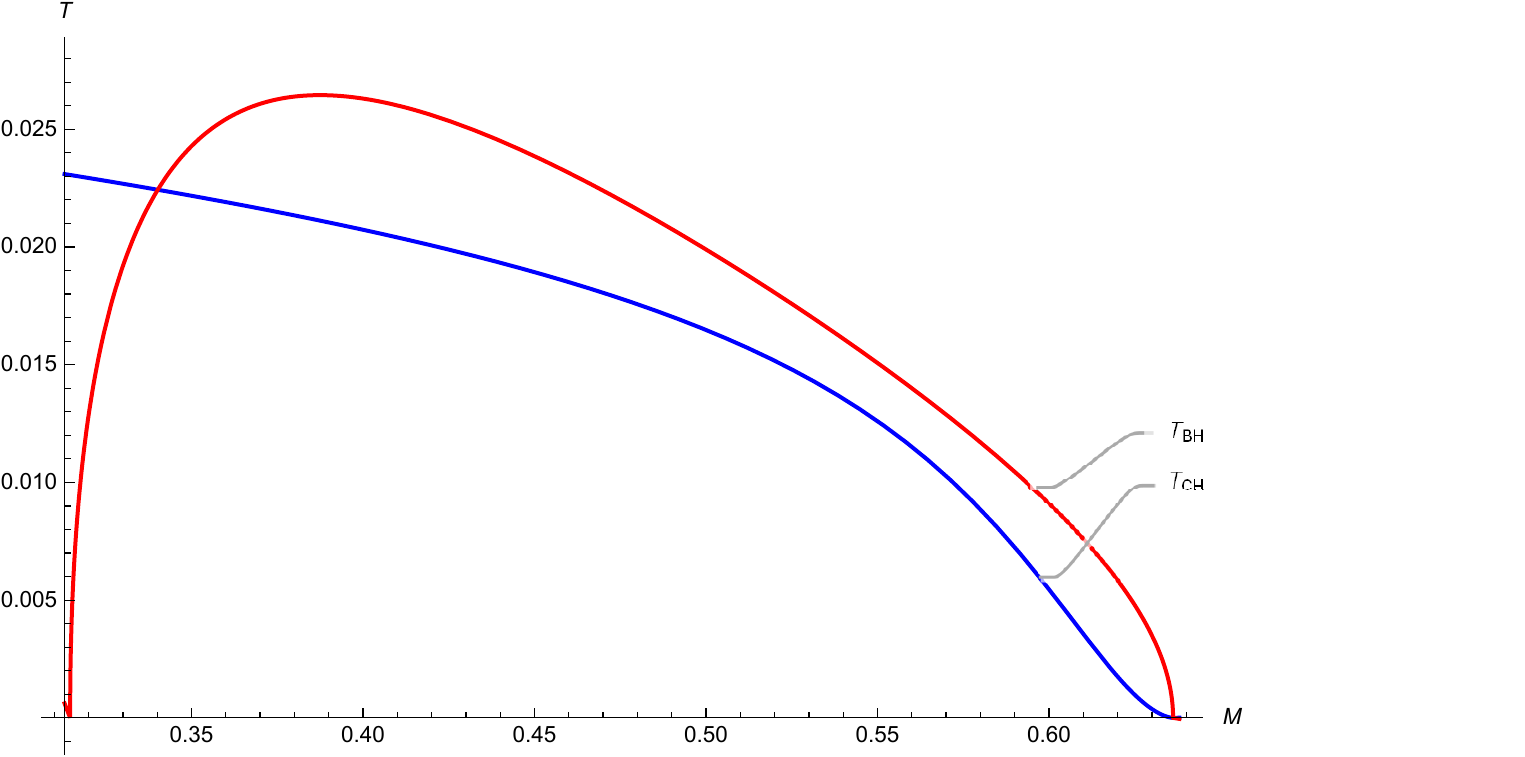}
		
		\includegraphics[height=4 cm,width=5.5 cm]  {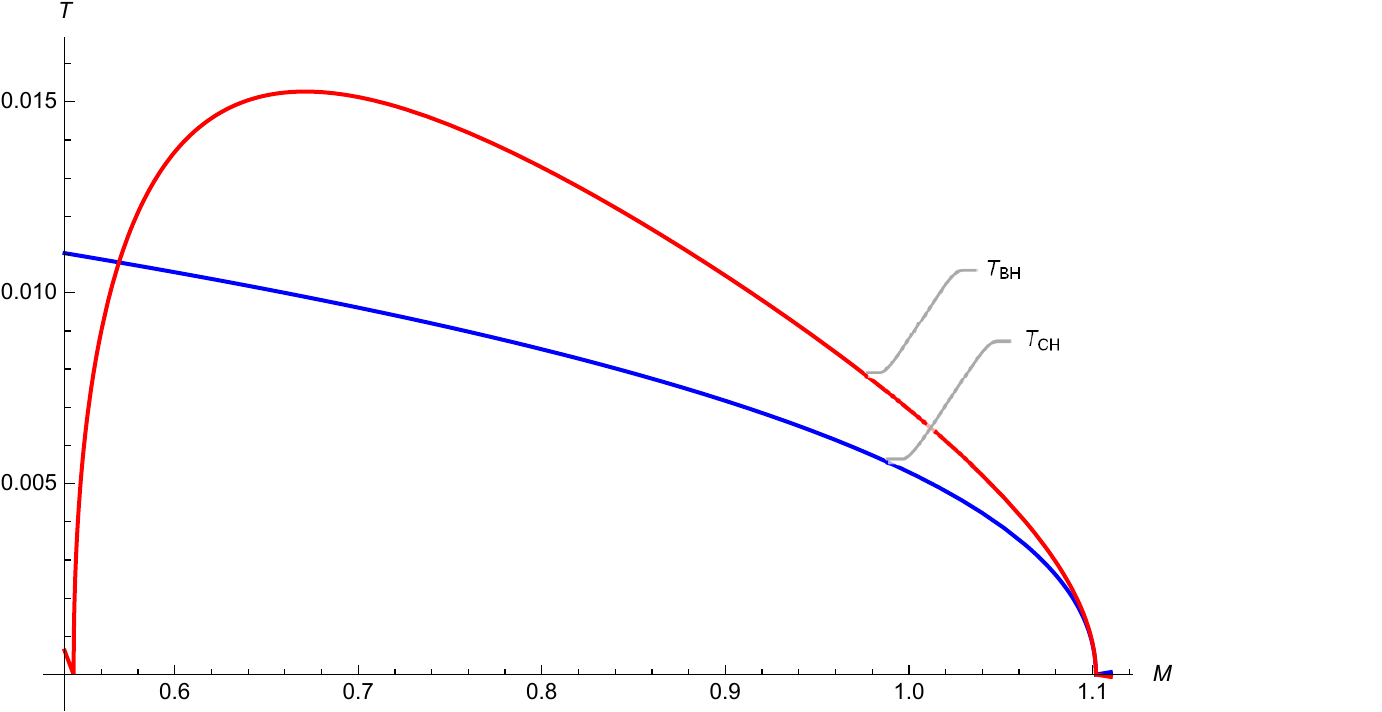}
		\hspace*{0.5cm}
		\includegraphics[height=4 cm,width=5.5 cm] {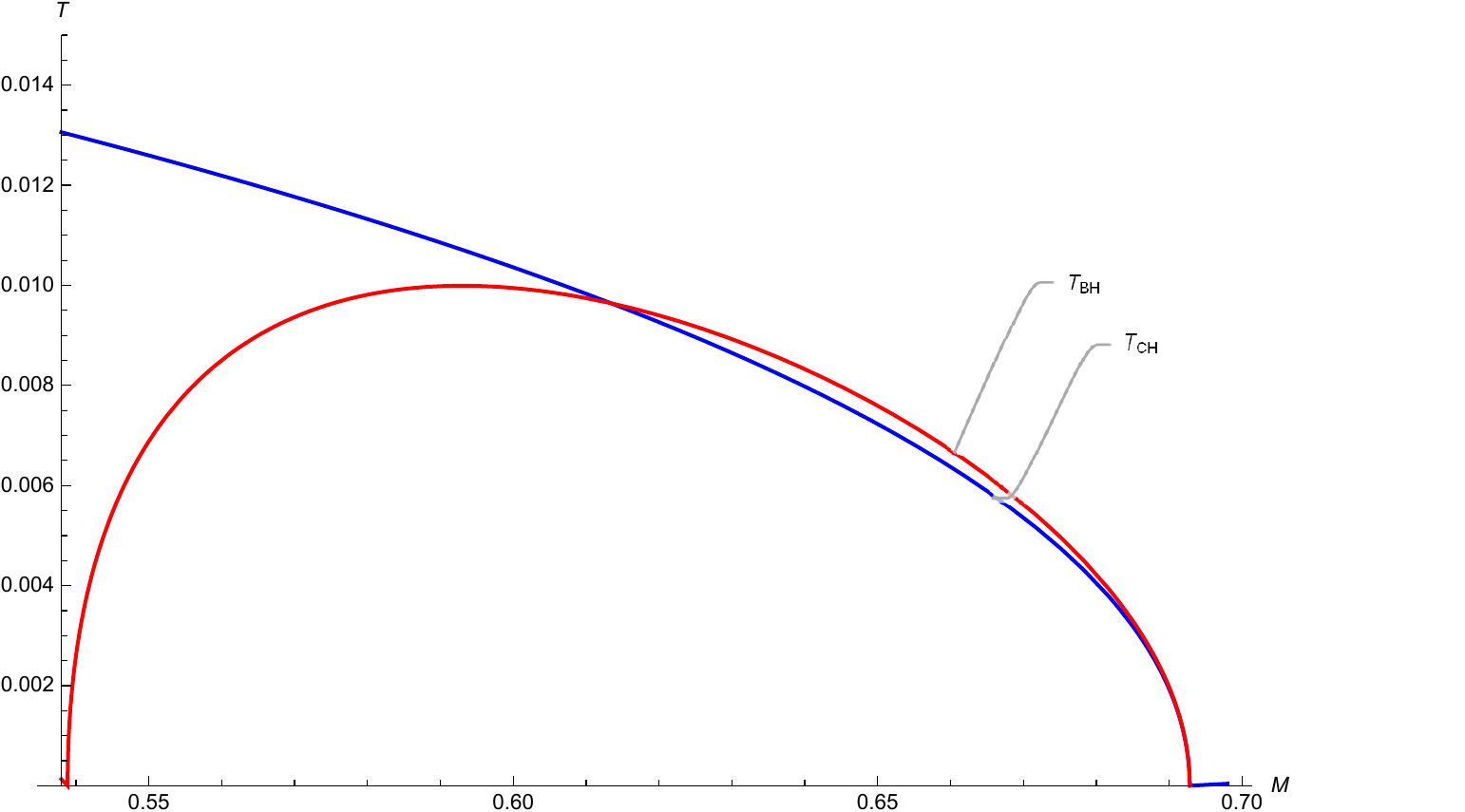}
		
		\caption{\scriptsize{Temperature of the black hole (Red line) and cosmological (Blue line) horizons for dS 4D EGB black hole versus the mass of the black hole. Plots have been depicted from top to bottom with $\alpha=0.1, 0.3$ and from left to right with $\Lambda=0.1,0.3$. Temperature behaviors for the dS 4D EGB black hole are an exact match with the temperature of a regular black hole in Lovelock gravity in higher dimensions.}}
		\label{figure_8}
	\end{figure}

	In the first step to study the particles' tunneling from the ds 4D EGB black hole, we need to find an analytical term for horizons from which tunneling calculations can be performed; the roots of Eq. (\ref{g00dsEGB}) will be as follows
	\begin{equation}\label{horizonds}
		r=\pm\frac{1}{2} \Big[\sqrt{\eta}\mp\sqrt{\frac{\eta-12 M}{\eta\Lambda}}\Big],
	\end{equation}
	where
	\begin{equation}
		\eta=\frac{2}{\Lambda}+\frac{3\times 2^{1/3}(1-4\alpha \Lambda)}{\Lambda \xi^{1/3}}+\frac{\xi^{1/3}}{3\times2^{1/3} \Lambda},
	\end{equation}
	and
	\begin{equation}
		\xi=972 \Lambda M^2 -648 \alpha \Lambda-54+\sqrt{(972 \Lambda M^2 -648 \alpha \Lambda-54)^2-4(9-36 \alpha \Lambda)^3}.
	\end{equation}
	Continuing with Eq. (\ref{horizonds}) as black hole horizons make it almost impossible to analyze horizons and perform tunneling calculations and obtain temperatures. For this reason, we will do the calculations of this section with the specified values for $\alpha$ and $\Lambda$ according to Fig. (\ref{figure_2}).\\
	If we take $\alpha=0.1$ and $\Lambda=0.1$, cosmological and black hole horizons become as follows
	
	\begin{equation}\label{horizonds11}
		r(M)=\pm\frac{1}{2}\Big[\sqrt{\eta}\pm\sqrt{40-\frac{6.33\times10^{18}}{\xi^{1/3}}-1.52\times10^{-17}\xi^{1/3}\mp\frac {120 M}{\sqrt{\eta}}}\Big],
	\end{equation}
	where
	\begin{equation}
		\eta=20+\frac{6.33\times10^{18}}{\xi^{1/3}}+1.52\times10^{-17}\xi^{1/3},
	\end{equation}
	and
	\begin{equation}
		\xi=10^{53}\Big(5.17 M^2+\sqrt{26.71 M^4-33.24 M^2 +3.05}-3.22\Big).
	\end{equation}
	As we illustrated in Fig. (\ref{figure_2}), in defined range of mass $M_{cri1}<M<M_{cri2}$, there are three horizons: $r_0$, $r_{BH}$ and $r_{CH}$. For example, if we take $M=0.7$ in Eq. (\ref{horizonds11}) then $r_0=0.075$, $r_{BH}=1.43$ and $r_{CH}=4.58$. In the following, we continue particles' tunneling calculation with considering two larger values of Eq. \ref{horizonds11} as a cosmological and black hole horizons as follows
	\begin{equation}\label{CHhorizonds11}
		r_{CH}(M)=\frac{1}{2}\Big[\sqrt{\eta}+\sqrt{40-\frac{6.33\times10^{18}}{\xi^{1/3}}-1.52\times10^{-17}\xi^{1/3}-\frac {120 M}{\sqrt{\eta}}}\Big],
	\end{equation} 
	and
	\begin{equation}\label{BHhorizonds11}
		r_{BH}(M)=\frac{1}{2}\Big[\sqrt{\eta}-\sqrt{40-\frac{6.33\times10^{18}}{\xi^{1/3}}-1.52\times10^{-17}\xi^{1/3}-\frac {120 M}{\sqrt{\eta}}}\Big].
	\end{equation}
	Note that the particle with energy $\omega$ tunnels from initial state in $r_{in}=r_{CH}(M)-\epsilon$ to final state in $r_{out}=r_{CH}(M-\omega)+\epsilon$. To calculation of imaginary part of action (Eq. (\ref{Imaction})), with considering outgoing geodesics, there are three poles in first integral; So, we expand the $\dot{r}$ in terms of $r_{out}$, outgoing geodesics becoms
	\begin{equation}\label{rdot2}
		\dot{r}=1-\sqrt{1-\Big(r-\frac{1}{2}(\sqrt{\delta}+\sqrt{\gamma})\Big)\Big(5 (1-\sqrt{1.01+\frac{6.4 (M-\omega)}{(\sqrt{\delta}\sqrt{\gamma})^3}})(\sqrt{\delta}+\sqrt{\gamma})+\frac{24 (M-\omega)}{(\sqrt{\delta}+\sqrt{\gamma})^2\Big[\sqrt{1.01+\frac{6.4 (M-\omega)}{(\sqrt{\delta}+\sqrt{\gamma})^3}}\Big]}\Big)},
	\end{equation}
	where
	\begin{equation}
		\delta=20-1.5  [435 (M-\omega)^2+147]^{1/3}-\frac{63}{[435 (M-\omega)^2+147]^{1/3}},
	\end{equation}
	and
	\begin{equation}
		\gamma=40+1.5  [435 (M-\omega)^2+147]^{1/3}+\frac{63}{[435 (M-\omega)^2+147]^{1/3}}-\frac{120 (M-\omega)}{\sqrt{\delta}}.
	\end{equation}
	Substituting Eq. (\ref{rdot2}) in Eq. (\ref{ImSgeneral}) and applying residue calculation, we solved the first integral of the imaginary part of the action. Considering that in the final answer, we need the coefficient of the first order of $\omega$ to gain the temperature of the cosmological horizon, to continue calculation, we expand the answer of the first integral in terms of $\omega$. Finally, by calculating the second integral and keeping the omega coefficient, we get the temperature of the cosmological horizon ($T_{CH}$) of the dS EGB black hole in terms of the mass of the black hole. To obtain the temperature of the black hole horizon ($T_{BH}$) of the dS EGB black hole, we do the same calculation by considering $r_{BH}$ (Eq. (\ref{BHhorizonds11})) as a particles' tunneling horizon. Also, we repeat the process to other values of the coupling constant $\alpha$ and the cosmological constant $\Lambda$. Eventually, we demonstrate the final result of all calculations in Fig. (\ref{figure_8}).\\
	The first important result that can be seen from the diagrams is the different behavior of temperature of the black hole and cosmological horizons. With reducing the mass of the black hole, the temperature of the black hole horizon goes up to reach a maximum temperature in the $M'$ and then it descends to zero when $M=M_{cri1}$; While the temperature of the cosmological horizon rises to the higher temperature until the mass of the black hole reaches to $M=M_{cri1}$. Also, the temperature of the cosmological horizon is less than the temperature of the black hole horizon ($T_{CH} < T_{BH}$) in the range of the mass ($M_{cri2}<M<M_{eq}$). This temperature behavior of the cosmological horizon has been seen in the dS Schwarzschild black hole \cite{Med02} and the Schwarzschild black hole surrounded by the quintessence \cite{EsNoz20}, But the temperature behavior of the black hole horizon is different from the result of Refs. \cite{Med02,EsNoz20}. Since, for the rang of the mass ($M_{eq}<M<M_{cri2}$), the temperature of the black hole horizon is larger than the temperature of the cosmological horizon ($T_{BH}>T_{CH}$), there is the heat flow from the region with high temperature to the region with low temperature. After that, when the mass of the black hole reaches to the $M_{eq}$, two temperature becomes equal together, $T_{BH}=T_{CH}$. On the other hand, when the mass of the black hole is in the range ($M_{cri1}<M<M_{eq}$), the evolution should be in the inverse direction. In other words, in this range of mass, the black hole horizon increases its temperature by absorbing radiation coming from the cosmological horizon. As a result, the mass of the black hole increases to the equal mass, $M_{eq}$. Finally, there is the stable rest mass with finite and equal temperature for the black hole and cosmological horizon.\\
	Besides, as shown in Fig. (\ref{figure_8}), increasing the coupling constant, $\alpha$, decreases both the black hole and cosmological temperature but it is more impressive to reducing the black hole temperature. On the other hand, increasing $\Lambda$ decreases the black hole temperature and increases the cosmological temperature. As a result, with increasing $\Lambda$, it has been predicted the larger equal mass $M_{eq}$ to equaling two temperatures.\\
	What is important and exciting for us is the similarity of these temperature behaviors of dS 4D EGB black hole with temperature behaviors of the n-fold degenerated GB black hole and pure Lovelock black hole \cite{ArEs19} in higher dimensions. Such a likeness in temperature behavior between these two black holes shows the deep connection between the spacetime of these two black holes. One has been derived in higher dimension Lovelock gravity and another has been derived in 4D EGB gravity.\\

	\section{Summery and Conclusion}
	In this work, we investigated the novel 4D EGB gravity in three asymptotical spacetimes and their thermodynamics, especially their radiation and temperature. First, we demonstrated the structure of 4D EGB spacetime with $\Lambda=0$ (asymptotically flat spacetime). In this case, the critical mass,$M_{cri}$, is associated with the coupling constant; For $M>M_{cri}$, the black hole has two horizons: the white hole horizon and the black hole horizon. Also, there is the minimum gravitational potential located at $r_0$ where the attractive and repulsive gravitational force is balanced. In such a spacetime, the existence of a coupling constant alone modifies the small radius regime. After, we calculated the emission rate and the correlation function between the radiated modes based on the method related to the tunneling of massless particles from the black hole horizon. We find the correlation between radiated modes and back-reaction effects. In the several references that we have introduced, it has been shown that the existence of a correlation between the radiated modes indicates unitary and solving the information loss problem automatically. Besides, we calculated the temperature of the black hole horizon of the 4D EGB black holes through the tunneling process. We illustrated that by reducing the mass, the temperature of the black hole increases to a maximum temperature and then decreases to lower temperatures. Finally, the rest mass with zero temperature remains. It is necessary to mention that increasing the coupling constant, $\alpha$, the maximum temperature and the rest mass of the black hole reduce. It is good to be mentioned that these behaviors of the temperature are exactly matched with the results of the calculation based on the standard definition of the Hawking temperature.\\
	At the second step, we glimpsed the asymptotically AdS 4D EGB spacetime. The existence of $\Lambda<0$ just affects the asymptotical behavior of the black hole and doesn't affect the number of horizons and general structure of the asymptotically flat 4D EGB spacetime. In this situation, when the mass of the black hole is greater than the critical mass, the black hole has two horizons. By plotting the mass of the black hole versus the radius of the black hole, we demonstrated decreasing $\alpha$ decreases the minimum mass (critical mass), and the changing $\Lambda$ affects the asymptotic behavior, as expected.\\
	At the final step, we explored the asymptotically dS 4D EGB spacetime. Such a spacetime in the specific range of the mass has three horizons: the cosmological horizon ($r_{CH}$), the black hole horizon ($r_{Bh}$), and the inner horizon ($r_{in}$). We illustrated the effects of the different coupling constant and cosmological constant on the dS 4D EGB gravity. According to $\alpha$ and $\Lambda$, these specific mass ranges could be defined as follows: $M=M_{cri1}$ defined where the inner and the black hole horizons merge, and $M=M_{cri2}$ defined where the black hole and the cosmological horizons merge. When the mass of the black hole be in the range ($M_{cri1}<M<M_{cri2}$), there exist three horizons. As we illustrated, increasing $\alpha$ and $\Lambda$ from $0.1$ to $0.7$, makes this specific mass range more limited. By selecting the proper $\alpha$ and $\Lambda$, we investigated the tunneling of massless particles from the black hole and the cosmological horizons. Clearly, with the study of the emission rate and calculating the correlation between the radiated mods, it can be seen that the correlation between modes and back-reaction effects, too. Finally, we calculated the temperature of the black hole horizon and the cosmological horizon separately. The temperature of the black hole horizon grows up to the maximum temperature and then it decreases to equal to the temperature of the cosmological horizon. On the other hand, the temperature of the cosmological horizon increases to reach the same as the temperature of the black hole horizon. We interpreted the evolution of these temperature horizons as follows: when the mass of the black hole reaches the critical mass ($M_{cri2}$), the temperature of the cosmological horizon increases with the absorbing the radiation from the black hole horizon. Besides, for consistency, the mass of the black hole decreases. On the other hand, in the inversed scenario, when the mass of the black hole reaches the $M=M_{cri1}$, the black hole horizon increases its temperature by absorbing the radiation from the cosmological horizon. In the process, the mass of the black hole increases. Finally, in this stage, the 4D EGB black hole has a stable rest mass with an equal finite temperature for both the black hole horizon and the cosmological horizon.


\end{document}